%% file: arXiv.tex
\documentclass{ETHpaper}

\usepackage{amsmath,amssymb,amsfonts}
\usepackage{graphicx}
\usepackage{textcomp}
\usepackage{xcolor}
\def\BibTeX{{\rm B\kern-.05em{\sc i\kern-.025em b}\kern-.08em
    T\kern-.1667em\lower.7ex\hbox{E}\kern-.125emX}}

\usepackage[numbers]{natbib}
\usepackage{tabularx}
\usepackage{booktabs}
\usepackage[capitalise]{cleveref}
\usepackage{multirow}
\usepackage{nicefrac}
\usepackage{csquotes}
\usepackage{tikz}
\usetikzlibrary{arrows}
\usetikzlibrary{arrows.meta}
\usetikzlibrary{backgrounds}
\usetikzlibrary{calc}
\usetikzlibrary{decorations.pathreplacing}
\usetikzlibrary{fit}
\usetikzlibrary{positioning}

\usepackage{algorithm}%
\usepackage{algorithmicx}%
\usepackage{algpseudocode}%
\usepackage{tcolorbox}
\usepackage[title]{appendix}

\definecolor{firstcolor}{RGB}{168,50,45} 
\definecolor{colorSG}{RGB}{168,50,45}
\definecolor{customY}{HTML}{FBB13C}
\definecolor{customG}{HTML}{218380}
\definecolor{customT}{HTML}{73D2DE}
\definecolor{customW}{HTML}{FCFCFF}
\definecolor{customB}{HTML}{2E5EAA}
\definecolor{customP}{HTML}{5B4E77}
\definecolor{colorDeveloper}{HTML}{F8766D}
\definecolor{colorDocumenter}{HTML}{A3A500}
\definecolor{colorProductOwner}{HTML}{00BF7D}
\definecolor{colorStakeholder}{HTML}{00B0F6}
\colorlet{colorInactive}{gray}
\definecolor{gitRedFaded}{HTML}{FFEEF0}
\definecolor{gitGreenFaded}{HTML}{E6FFED}
\definecolor{gitRed}{HTML}{FFDCE0}
\definecolor{gitGreen}{HTML}{CDFFD8}
\definecolor{gitRedFull}{HTML}{CB2431}
\definecolor{gitGreenFull}{HTML}{2CBE4E}
\definecolor{net_2005}{RGB}{163, 159, 148}
\definecolor{net_2015}{RGB}{175, 153, 124}
\definecolor{net_both}{RGB}{108, 99, 86}

\colorlet{colorScenarioObserved}{customB}
\colorlet{colorScenarioAgile}{colorSG!40}
\colorlet{colorScenarioRoles}{gray!40}

\newcommand{\code}[1]{\textsc{\detokenize{#1}}}

\newcommand{\developer}{\textsc{Developer}}

\newcommand{\documenter}{\textsc{Documenter}}

\newcommand{\stakeholder}{\textsc{Stakeholder}}

\newcommand{\productowner}{\textsc{Product Owner}}

\tikzstyle{legendbullet}=[circle, inner sep=0pt, minimum size=2mm]
\newcommand{\genuaRoleLegend}[1]{
    \begin{tikzpicture}
        \node[legendbullet, fill=colorDeveloper] (bulletDeveloper) at (0, 0) {};
        \node[right=0mm of bulletDeveloper] (labelDeveloper) {\footnotesize \developer{} \ifthenelse{\equal{#1}{1}}{(\code{Dev})}{}};

        \node[legendbullet, fill=colorDocumenter, below=2mm of bulletDeveloper] (bulletDocumenter) {};
        \node[right=0mm of bulletDocumenter] (labelDocumenter) {\footnotesize \documenter{} \ifthenelse{\equal{#1}{1}}{(\code{Doc})}{}};

        \node[legendbullet, fill=colorProductOwner, right=8mm of labelDeveloper] (bulletProductOwner) {};
        \node[right=0mm of bulletProductOwner] (labelProductOwner) {\footnotesize \productowner{} \ifthenelse{\equal{#1}{1}}{(\code{PO})}{}};

        \node[legendbullet, fill=colorStakeholder, below=2mm of bulletProductOwner] (bulletStakeholder) {};
        \node[right=0mm of bulletStakeholder] (labelStakeholder) {\footnotesize \stakeholder{} \ifthenelse{\equal{#1}{1}}{(\code{Sta})}{}};

        \node[anchor=east, left=6mm of $(bulletDeveloper.west)!0.5!(bulletDocumenter.west)$] {\textbf{Role}};
    \end{tikzpicture}
}

\makeatletter
\renewenvironment*{displayquote}
  {\begingroup\setlength{\leftmargini}{.5cm}\csq@getcargs{\csq@bdquote{}{}}}
  {\csq@edquote\endgroup}
\makeatother

\begin{document}

\title{Locating Community Smells in Software Development Processes\\ Using Higher-Order Network Centralities}
\titlealternative{Locating Community Smells in Software Development Processes Using Higher-Order Network Centralities}

\makeatletter
\newcommand{\linebreakand}{%
  \end{@IEEEauthorhalign}
  \hfill\mbox{}\par
  \mbox{}\hfill\begin{@IEEEauthorhalign}
}
\makeatother

\author{Christoph Gote\textsuperscript{1,3,4,6} \qquad Vincenzo Perri\textsuperscript{3,4,7} \qquad Christian Zingg\textsuperscript{1,8} \qquad Giona Casiraghi\textsuperscript{1,9} \\ Carsten Arzig\textsuperscript{2,10} \qquad Alexander von Gernler\textsuperscript{2,11} \qquad Frank Schweitzer\textsuperscript{1,5,12} \qquad Ingo Scholtes\textsuperscript{3,4,13}}
\authoralternative{C. Gote, V. Perri, C. Zingg, G. Casiraghi, C. Arzig, A. von Gernler, F. Schweitzer, and I. Scholtes}
\address{
\textsuperscript{1}Chair of Systems Design, ETH Zurich, Zurich, Switzerland \\
\textsuperscript{2}genua GmbH, Kirchheim bei München, Germany\\
\textsuperscript{3}Chair of Machine Learning for Complex Networks, Center for Artificial Intelligence and Data Science (CAIDAS), University of Würzburg, Würzburg, Germany \\
\textsuperscript{4}Data Analytics Group, Department of Informatics, University of Zurich, Zurich, Switzerland \\
\textsuperscript{5}Complexity Science Hub, Vienna, Austria\\[2mm]
\textsuperscript{6}corresponding author: cgote@ethz.ch \quad
\textsuperscript{7}vincenzo.perri@uni-wuerzburg.de \quad
\textsuperscript{8}czingg@ethz.ch \quad
\textsuperscript{9}gcasiraghi@ethz.ch \\
\textsuperscript{10}carsten\_arzig@genua.de \quad
\textsuperscript{11}alexander\_gernler@genua.de \quad
\textsuperscript{12}fschweitzer@ethz.ch \quad
\textsuperscript{13}ingo.scholtes@uni-wuerzburg.de
}

\maketitle

\begin{abstract}
Community smells are negative patterns in software development teams' interactions that impede their ability to successfully create software.
Examples are team members working in isolation, lack of communication and collaboration across departments or sub-teams, or areas of the codebase where only a few team members can work on.
Current approaches aim to detect community smells by analysing \emph{static} network representations of software teams' interaction structures.
In doing so, they are insufficient to locate community smells within development \emph{processes}.
Extending beyond the capabilities of traditional social network analysis, we show that higher-order network models provide a robust means of revealing such hidden patterns and complex relationships.
To this end, we develop a set of centrality measures based on the \texttt{MOGen} higher-order network model and show their effectiveness in predicting influential nodes using five empirical datasets.
We then employ these measures for a comprehensive analysis of a product team at the German IT security company \textit{genua GmbH}, showcasing our method's success in identifying and locating community smells.
Specifically, we uncover critical community smells in two areas of the team's development process.
Semi-structured interviews with five team members validate our findings:
while the team was aware of one community smell and employed measures to address it, it was not aware of the second.
This highlights the potential of our approach as a robust tool for identifying and addressing community smells in software development teams.
More generally, our work contributes to the social network analysis field with a powerful set of higher-order network centralities that effectively capture community dynamics and indirect relationships.
\end{abstract}

\section{Introduction}\label{sec:introduction}

We are woken up by the alarm app on our phones; we use navigation software to commute to work; we use social media to interact with friends and collaborators around the world.
These seemingly trivial everyday activities demonstrate our reliance on software and the resulting importance of its effective creation and maintenance for our modern society.

Software is typically created by teams.
When joining a team, developers have different specialisations and past experiences.
Thus, all developers have personal knowledge unique to them.
For example, team members could be familiar with different programming languages, have expertise in frontend or backend development, or have an elaborate knowledge of a product's past design decisions.
The key to a software team's success is the effective combination of the unique knowledge of its members \citep{blackler1995knowledge}.

When team members leave their team, they take their unique knowledge with them.
Therefore, if knowledge is not adequately shared and managed, this unique knowledge is  lost to the remaining team.
In software development, such perils are referred to as \emph{social debt}.
Social debt is defined as the ``unforeseen project cost connected to a `suboptimal' development community'' \citep{tamburri2013social}.
The sources of social debt within a team are called \textit{community smells} \citep{tamburri2013social,caballero2022community}.

Community smells can only be addressed once they have been identified and understood.
To this end, Social Network Analysis (SNA) on static networks, in which nodes represent developers and edges dyadic interactions between them, are a widely used tool~\citep{meneely2008predicting,bird2009putting,tamburri2015social,tamburri2019exploring,tamburri2019software,almarimi2020learning}.
However, \emph{static} network models only allow insights into \emph{direct} relations between team members.
Therefore, they cannot consider patterns in development \emph{processes}.
As we show in this paper, this means they can only be used to study patterns in the team's direct interaction \emph{topology} and, thus, that trivial things such as bottlenecks in software development \emph{processes}---i.e., in the sequences of actions team members need to perform to create software---would be completely overlooked.

To capture patterns in the \textit{dynamics} between team members, \textit{higher-order} generalisations of network models have been proposed \citep{lambiotte2019networks,battiston2020networks,torres2020and}.
While the specific assumptions about the higher-order patterns captured by those models differ, they have in common that they generalise network models towards representations that go beyond pairwise, dyadic interactions.

This paper explores the use of higher-order network analysis to identify community smells in the development processes of software teams.
Specifically, we analyse the development activities of a product team at the German IT security company \textit{genua GmbH} using centrality measures computed on the higher-order network model \texttt{MOGen} \citep{gote2020predicting}, which captures non-Markovian patterns in paths in complex networks, i.e., patterns that require memory to be modelled.
Our contributions are as follows:
\begin{itemize}
    \item We mine a unique curated dataset comprised of fine-grained time-stamped path data from two issue trackers and a code review platform tracking the actions of a product team at \textit{genua} over 20 years.
    \item To identify community smells, we consider five centrality measures that serve as proxy for the influence of specific nodes and node sequences in dynamical processes.
    For these centrality measures, we demonstrate that utilising a \emph{path model} results in improved predictions of influential nodes in time-series data compared to a simpler \emph{network-based model}, provided there is enough training data. However, this approach results in a significant generalisation error for smaller datasets.
    \item To address this problem, we define equivalent measures for \texttt{MOGen}, a higher-order generative model for paths in complex networks \citep{gote2020predicting}. We show that our \texttt{MOGen}-based centrality measures effectively mitigate the generalisation error, i.e., they balance between underfitting and overfitting the data.
    \item We apply our five \texttt{MOGen}-based centrality measures to identify team members consistently taking over tasks that no other team members perform, identifying two community smells in the \textit{dynamic} development process that could have not been identified using static SNA.
    \item We validate our findings in semi-structured interviews with 5 developers from \textit{genua}.
    The team is aware of one of these community smells and employs active measures against it.
    However, the team was not aware of the second community smell but could confirm it ex-post.
    Thus, we prove that our approach successfully uncovers community smells and can aid software teams in countering them.
\end{itemize}

This article is an extended version of the ASONAM 2022 contribution “Predicting Influential Higher-Order Patterns in Temporal Network Data” \citep{gote2022predicting}. 
In this version of our work, we extend our previous development and assessment of \texttt{MOGen}-based centrality measures by applying them to the empirical software engineering domain (\Cref{sec:data,sec:genua_analysis,sec:genua_interviews}).
In doing so, we show the capability of higher-order network methods in a real-world application.

\section{Related work}\label{sec:related_works}

\subsection{Community smells}
Social debt in software development refers to the ``unforeseen project cost connected to a `suboptimal' development community'' \citep{tamburri2013social}.
The sources of social debt are called community smells, which have been identified as a cause of issues in the source code \citep{palomba2017social, palomba2018beyond}, e.g., using them to predict bugs \citep{eken2021empirical}.

The empirical software engineering literature has identified a wide variety of community smells~(see \citealp{caballero2022community} for a recent review).
Community smells are usually caused by a lack of communication and knowledge exchange between individuals or subgroups within the team \citep{lin2017}.
This can be due to highly independent development tasks, in which developers work without communicating with others.
Such developers are referred to as ``lone wolves'' \citep{tamburri2019exploring}.
Similarly, groups of developers not communicating with the remaining team constitute ``organisational silos'' \citep{rilling2008beyond}.
Over time, this lack of communication can create a large ``cognitive distance'' between team members, obfuscating and impeding interactions (the ``black cloud effect'').
Ultimately, such problems result in a further lack of interactions, referred to as ``bottlenecks'' or ``radio silence'' \citep{tamburri2015social}.
Additionally, a large cognitive distance between team members can promote the emergence of other community smells, such as ``prima donnas'', ``sharing villainy'', or ``code-red'' situations, which adversely impact development and reduce overall trust within the team \citep{tamburri2015social}.
Here, prima donnas are developers who believe that their work is superior to that of their colleagues and are unwilling to collaborate or share knowledge with others.
Similarly, sharing villainy is the tendency of some developers to hoard knowledge and resources, making it difficult for others to access and use them \citep{tamburri2015social}.
Finally, \textit{code-red} \citep{palomba2018beyond} refers to situations where only a few developers are capable of maintaining certain areas of the codebase.

Several studies have shown that community smells can have a significant negative impact on software development projects, including delays, lower productivity, and increased technical debt \citep{sedano2017,ma2020}.
For example, community smells can predict if and which developers stop contributing to a project \citep{huang2021predicting,huang2022community}.
The consequences are particularly severe if the departing developers were lone wolves, part of an organisational silo, or maintainers of code-red code. This is expressed in the ``bus number'' or ``truck factor'' which counts the number of developers that have to be `hit by a bus', before a software development project would come to a halt \citep{izquierdo2009using,avelino2016novel,cosentino2015assessing,ricca2011difficulty,ferreira2016comparative,ferreira2019algorithms}.
Through the dependency network, community smells can further be amplified resulting in a negative impact across the entire software ecosystem \citep{schueller2022evolving,schueller2022modeling}.

Once community smells have been identified, measures can be employed to counter them.
For example, if the community smell is due to insufficient knowledge distribution, the team can enhance task assignment \citep{etemadi2022} through the use of agile methodologies such as Scrum \citep{Schwaber2020} or Extreme Programming \citep{Beck1999}.
Moreover, effective communication strategies such as regular team meetings, code reviews, and pair programming can help to reduce cognitive distance and promote knowledge sharing within the team \citep{lin2017}.
In summary, the identification and mitigation of community smells are crucial for managing social debt in software development projects.

\subsection{Social network analysis}

Social Network Analysis (SNA) is an established method to study community smells.
These works typically study \emph{static} networks, where nodes represent developers and edges dyadic interactions between them.
For example, \citet{meneely2008predicting} investigated collaboration structures using developer networks derived from code churn information to predict file-level failures.
Similarly, \citet{bird2009putting} employed SNA to study coordination among groups of developers with socio-technical dependencies and demonstrated that network properties of a software component could predict fault-proneness more accurately than dependency or contribution information alone.
\citet{tamburri2019exploring} developed an automated approach to identify community smell types based on SNA.
In related work, \citet{tamburri2019software} suggested that studying social debt and community smells at the architecture level could help software development communities eliminate critical organisational flaws and reduce costs.
\citet{almarimi2020learning} showed that network centralities were among the most influential characteristics identifying community smells. 

Conventional SNA methods are highly effective in identifying patterns within \emph{direct} relationships captured in network topology.
However, in most networked systems with sparse interaction topologies, the actual complexity stems from higher-order patterns representing \emph{indirect} influences~\citep{lambiotte2019networks}, which cannot be explained by network topology alone.
To account for these patterns, higher-order generalizations of network models have been proposed.
These higher-order models capture information about the network structure beyond pairwise, dyadic interactions.
Researchers have used higher-order methods to model memory in paths on networks.
For example, when modelling itineraries through a transportation system we need to not only consider where passengers are and where they can go (i.e., the transportation network), but also where they came from (i.e., memory).
To study random walks and diffusion processes \citep{scholtes2014causality,rosvall2014memory,lambiotte2015effect}, detect communities and assess node centralities \citep{rosvall2014memory,scholtes2016higher,xu2016representing,edler2017mapping,peixoto2017modelling}, analyze memory effects in clinical time-series data \citep{palla2018complex,krieg2020higher,myall2021network}, generate node embeddings and network visualizations based on temporal network data \citep{saebi2020honem,tao2017honvis,perri2020}, detect anomalies in time-series data on networks \citep{saebi2020efficient,larock2020hypa}, enhance deep learning models for networks \citep{qarkaxhija2022bruijn}, or assess the controllability of networked systems \citep{zhang2021higher}.

Furthermore, recent research has demonstrated the advantages of \emph{multi-order models}, which combine multiple higher-order models, for various applications such as the generalisation of PageRank to time-series data \citep{scholtes2017network} and path prediction in networks \citep{gote2020predicting}.
By accounting for paths of different lengths, multi-order models offer a more comprehensive understanding of complex networked systems than higher-order models alone~\citep{gote2020predicting}.
\citet{gote2022predicting} introduce centrality measures based on multi-order models.
In this paper, we show that such centrality measures enable a deeper analysis of team interactions, information flow, and knowledge-sharing patterns.
In the context of software development processes, this means that we can better identify potential bottlenecks, organisational silos, and lone wolves, thus enhancing our ability to detect and address community smells.

\section{Data}\label{sec:data}

To develop and test an approach to detect community smells in software development processes, we require fine-grained temporal data.
In particular, the ideal dataset contains information on which team member performed which action at which point in time.
To obtain such data, we collaborated with \textit{genua GmbH}, a German IT security company.
As part of this collaboration, \textit{genua} gave us full access to the development repository, the corresponding issue trackers, and the code review platform for one of their core projects with a development history of more than 20 years.
In addition, we had the opportunity for several extended discussions and interviews with multiple members of the development team.

\subsection{The development process at genua}\label{sec:genua_development_process}

\begin{figure}[t!]
	\centering
	\includegraphics[width=\columnwidth]{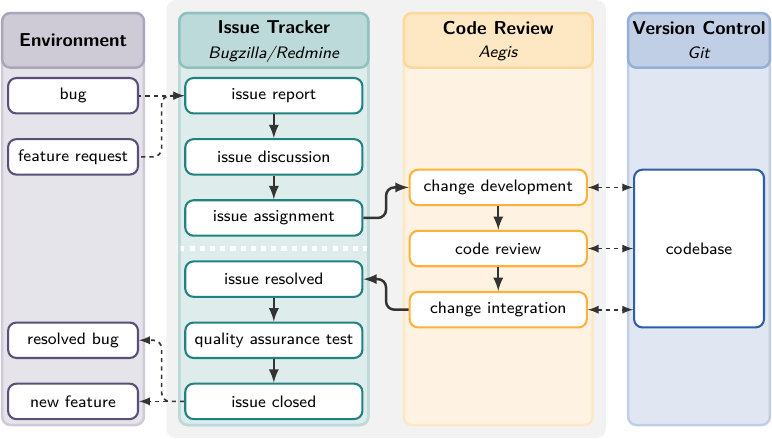}
	\caption{Simplified representation of the process to resolve a typical issue at \textit{genua}. The process takes place over multiple platforms. Backward loops have been removed.}
	\label{fig:Genua_Development_Process}
\end{figure}

Based on the availability of extensive and long-running data, we focused our data collection efforts on a single team developing one of \textit{genua}'s core products over the past 20 years.
In the following, we summarise the team's development process, which we visualise in \Cref{fig:Genua_Development_Process}.
The development team differentiates between bugs and features as two fundamental issue categories.
However, while the prioritisation process differs, the development process for features and bugs is identical.
Therefore, in the following, we refer to both as issues.

Whenever a team member finds a new bug or identifies a need for a new feature, an issue is created on the issue tracker.
Issues can be created by anyone with a technical or development-related position.
However, most new issues are created either by customer managers reporting bugs or feature requests from clients using \textit{genua}'s products or by developers identifying new bugs while working on the codebase.

In weekly bug meetings and daily Scrum meetings, the team discusses the prioritisation of currently open issues.
Subsequently, potential issue resolutions are discussed primarily on the issue tracker.
Once a promising issue resolution has been identified, the issue is assigned to a team member.
Similar to OSS projects \citep{crowston2007self}, this commonly occurs through self-assignment.

After being assigned the issue, the team member starts developing the corresponding change on the team's code review platform.
As an IT security company, one of \textit{genua}'s primary aims is to ship code with as few bugs as possible.
Therefore, each developed change is reviewed by a second team member and integrated into the release version of the codebase by a third team member.
Following this process, each change has been worked on by at least three separate team members.
Our data shows that for around 60\% of all issues, development, review, and integration are successfully completed in direct succession.
However, for the remaining issues, either the code review or the integration fails, requiring further development, thus repeating the process.
For readability, the resulting backward loops are not shown in \Cref{fig:Genua_Development_Process}.
However, we note that all completed changes start with the development of a change and end with its integration.

After successfully integrating a change, the corresponding issue is marked as resolved on the issue tracker.
Before closing the issue, final quality assurance tests are performed to ensure that the bug or feature works as intended in a wide range of application scenarios.
If this test is successful, the issue is closed.
Otherwise, the process continues with further discussion or development.

\subsection{Extracting paths capturing the development process}\label{sec:path_extraction_genua}

Until 2010, the team used \textit{Bugzilla} \citep{mozilla1998bugzilla} to track their issues.
From then onwards, the team transitioned to \textit{Redmine} \citep{lang2006redmine}.
For code review, the team uses the tool \textit{Aegis} \citep{miller2013aegis}.
To obtain all actions performed by team members for all issues and changes, we developed crawlers processing the internal databases and data storage structures of \textit{Bugzilla}, \textit{Redmine}, and \textit{Aegis}.
\textit{Bugzilla}'s and \textit{Redmine}'s backends use a MySQL database.
To obtain time-stamped records of team members' actions, labeled by issue, we accessed these databases and selected the respective columns.
\textit{Aegis} instead uses a tree of text files as internal data storage.
Here, we use a simple text-parsing tool that extracts the issues and the developer actions from these files.
We performed name disambiguation to uniquely identify each team member across all platforms.
Finally, we used a text-based approach to match the issues between the three platforms.
All team members and issues are represented by pseudo-anonymised IDs.
For privacy and security reasons, we were not allowed to process any user information or the content of team members' actions.

\begin{figure}
	\centering
	\includegraphics[width=.8\columnwidth]{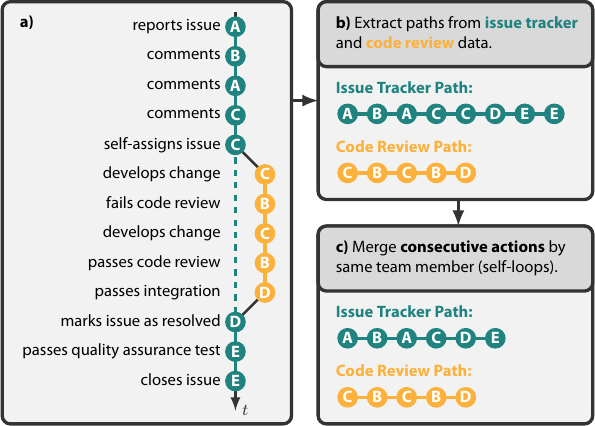}
	\caption[Process of extracting issue tracker and code review paths for an issue.]{Process of extracting issue tracker and code review paths for an issue. Nodes on paths represent actions by different team members $A$ to $E$. The required consecutive actions to start and end development, review, and integration in the code review process have been merged into single steps.}
	\label{fig:Path_Extraction}
\end{figure}

We capture the development process for each issue as a time-resolved path, sequentially connecting all team members contributing to it.
To obtain paths, we order all actions according to their time-stamp for each issue in ascending order.
We show an example of a resulting path in \Cref{fig:Path_Extraction}a.
Here, team member $A$ reports an issue and engages in a discussion with $B$ and $C$, each leaving a comment.
After the discussion, $C$ self-assigns the issue and starts developing the change required to resolve the issue.
Once $C$ completes the development process, $B$ reviews the changes.
However, the review fails as the code still contains errors.
After $C$ fixes the errors in a renewed development process, $B$ passes the code review, and $D$ integrates the changes into the main development branch.
Once integrated, team member $D$ marks the issue as resolved on the issue tracker, and $E$ performs a quality assurance test before closing the issue.

We split the path into separate paths for issue tracker (\textit{Bugzilla} and \textit{Redmine}) and code review (\textit{Aegis}) platforms (cf. \Cref{fig:Path_Extraction}b).
This is done for three reasons:
First, while the development, review, and integration of each change must be performed by three different team members, no such constraint exists for issue tracker platforms.
Second, the code review platform is only used by team members with development responsibilities---a subset of all team members.
Third, we can expect that team members adapt their behaviour to the platform they work on, which could lead to different behaviours of team members on the two platforms, which should be distinguished \citep{lisiecka2016medium}.
In total, we extract a combined total of 18,949 paths from \textit{Bugzilla} and \textit{Redmine}, and 24,360 paths from \textit{Aegis}\footnote{Note that some issues take multiple changes to resolve, explaining the higher path count for \textit{Aegis}.}.

\subsection{Characteristics of the software development team}\label{sec:data_genua_characteristics}
\begin{figure}[t!]
    \centering
    \includegraphics[width=.8\columnwidth]{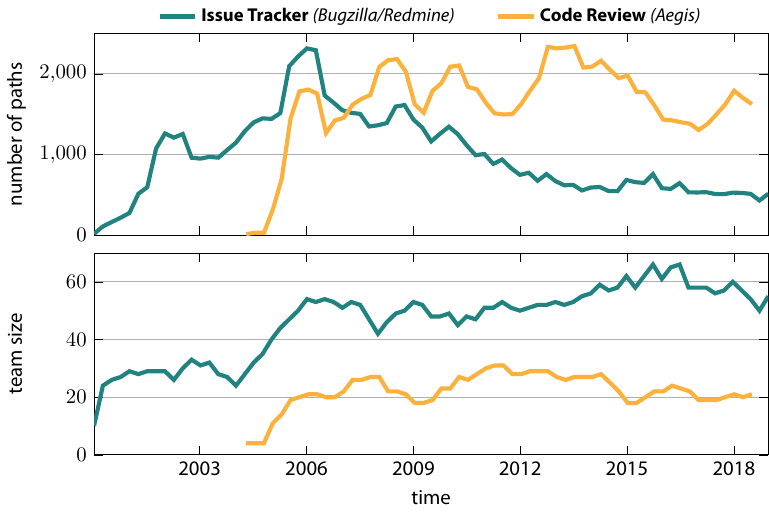}
    \caption[Path counts and team size for the issue tracker and code review data over time.]{Path counts and team size for the issue tracker and code review data over time. Counts were obtained for a rolling one-year window shifted by three-month increments.}
    \label{fig:BugzillaRedmine_Aegis_Number_Of_Paths_comb}
\end{figure}

\begin{figure}[t!]
    \centering
    \includegraphics[width=.85\textwidth]{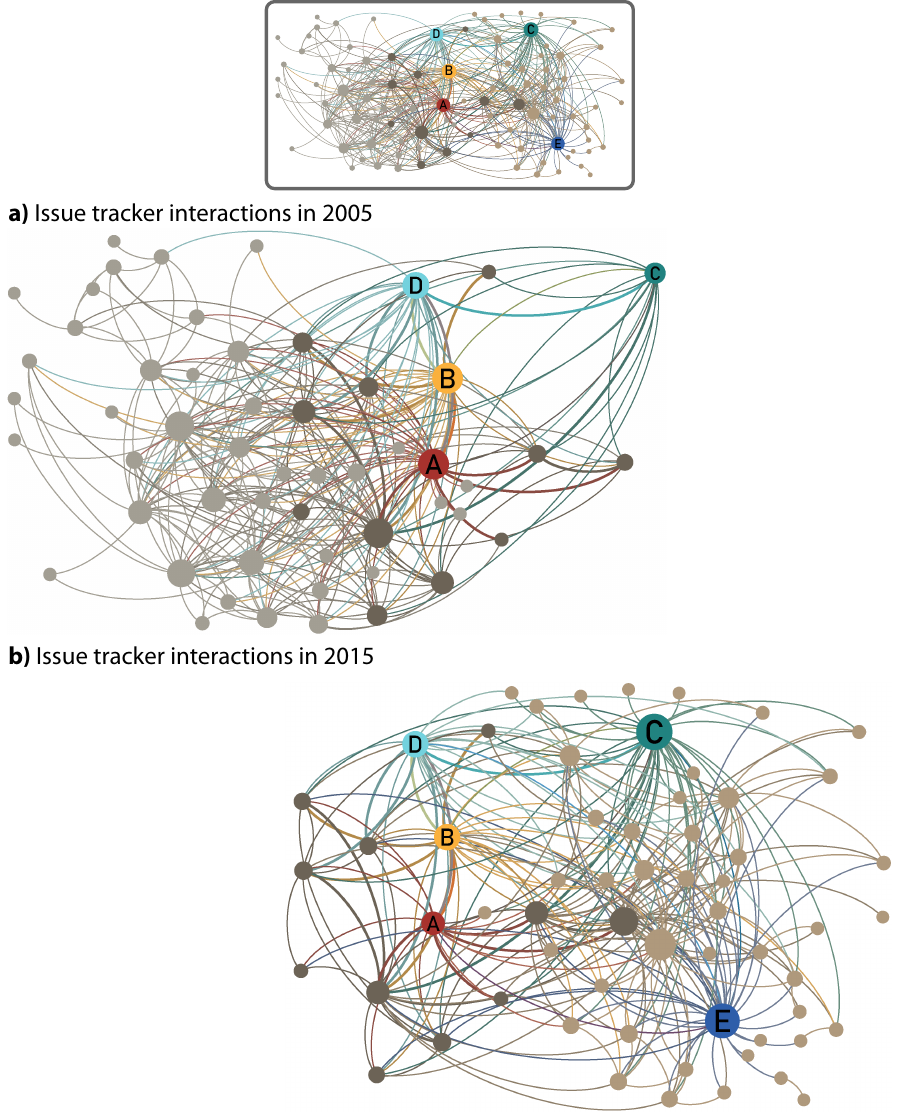}
    \caption{Issue tracker interaction networks for (a) 2005  and (b) 2015. The network layout is identical for both networks, i.e., all nodes appear in the same positions in (a) and (b). The full layout is shown above. The node sizes correspond to the number of interaction partners in the given year. Five team members $A$--$E$ that we analyse in detail in \Cref{sec:genua_analysis} are highlighted. All team members that appear in both 2005 and 2015 are shown in \textcolor{net_both}{\faSquare}. The team members only appearing in 2005 and 2015 are displayed in \textcolor{net_2005}{\faSquare} and \textcolor{net_2015}{\faSquare}, respectively.}
    \label{fig:genua_network_growth}
\end{figure}

In \Cref{fig:BugzillaRedmine_Aegis_Number_Of_Paths_comb}, we show the extracted path counts for a one-year rolling window shifted by three-month increments.
Apart from the very beginning of our data, we have at least 500 paths for each platform in any given one-year interval.
In the issue tracker data, we observe a peak at around 2,500 paths per year in 2006.
Afterwards, the number gradually declines until stabilising at around 500 paths per year.
With path counts averaging 1,800, the number of code review paths are significantly more consistent over time.

\Cref{fig:BugzillaRedmine_Aegis_Number_Of_Paths_comb} also shows the number of team members involved in the issue tracking and code review processes.
Only those team members directly involved with the development of changes to resolve issues interact on \textit{Aegis}.
Therefore, with 20 to 30 team members, this \textit{core team} is significantly smaller than the \textit{full team} interacting on the issue trackers, which also includes customer managers, the developers responsible for quality assurance, and members of other product teams.
While the size of the full team grew significantly from around 25 team members in 2000 to around 60 team members in 2018, the size of the core team has been relatively stable over time.

In \Cref{fig:genua_network_growth}, we show interaction networks derived from all issue tracker paths in 2005 and 2015, respectively.
To allow a comparison of the two teams, we fixed the network layout between the two years.
For reference, \Cref{fig:genua_network_growth} also shows the network derived from all paths occurring in either 2005 or 2015.
The node sizes correspond to the node degree---i.e., the number of interaction partners---of each team member.
The figure highlights that the team has changed substantially within the ten years, from 2005 to 2015.
Only around 25\% of the team members present in 2005 are still part of the team in 2015.
Instead, we see a large number of new team members in 2015.
In addition, the activity of team members has changed significantly.
In 2005, team members $A$ and $B$ are among those with the highest number of interaction partners.
In contrast, $C$, with a relatively small degree, is in the network's periphery.
By 2015, the roles are reversed, with $C$ and $E$---who was not even a part of the team in 2005---being the most connected nodes in the network, whereas the degrees of $A$ and $B$ have shrunk.
Overall, 176 unique people contribute to either the issue tracker or the code review platforms during our 20-year observation period.

\section{Methods}\label{sec:methods}

In the following, we introduce the methods used throughout this manuscript to analyse \textit{genua}'s development process.
Specifically, in \Cref{sec:paths_on_network_topologies}, we first introduce some terminology formally defining networks and paths.
In \Cref{sec:modelling_higher_order_patterns}, we motivate the use of higher-order models before introducing the multi-order model \texttt{MOGen} \citep{gote2020predicting} in \Cref{sec:MOGen}.
Finally, we introduce the \texttt{MOGen}-based centrality measures predicting influential nodes and higher-order patterns to ultimately detect community smells in \Cref{sec:centrality_measures}.

\subsection{Paths on network topologies}\label{sec:paths_on_network_topologies}

We mathematically define a \textit{network} as a tuple $G = (V,E)$, where $V$ is a set of nodes and $E$ is a set of edges.
As a real-world example, let us consider a public transport system.
Here, the individual stations are the nodes, and an edge exists between two nodes if there is a direct connection between the two stations.
A user of the system moves from a start to a destination following a \textit{path} that is restricted by the network topology.
A \textit{path} is defined as an ordered sequence $s = v_1 \rightarrow v_2 \rightarrow \dots \rightarrow v_{l_s}$ of nodes $v_i \in V$, where $l_s$ is the length of the path, and nodes can appear more than once.
We refer to a set of paths constrained by the same network topology as path dataset $P$.

While empirical paths can come from various sources, we can differentiate between two main types: (i) data directly recorded in the form of paths; (ii) paths extracted from data on temporal interactions, i.e., a temporal network.
Examples for the first case include clickstreams of users on the Web or data capturing passenger itineraries from public transport systems.
The primary example of temporal data are records on human interactions, which are a common source for studying knowledge transfer or disease transmission.

A \textit{temporal network} is a tuple $G^{(t)}=(V,E^{(t)})$, where $V$ is a set of vertices and $E^{(t)}$ is a set of edges with a time-stamp $E^{(t)} \subseteq V \times V \times \mathbb{N}$.
We can extract paths from a temporal network by setting two conditions. 
First, for two temporal edges $e_i = (v_1,v_2;t_1)$ and $e_j = (v_2,v_3;t_2)$ to be considered consecutive in a path---i.e., $s = \cdots \rightarrow v_1 \rightarrow v_2 \rightarrow v_3 \rightarrow \cdots$---they have to respect the arrow of time, i.e., $t_1 < t_2$.
Second, consecutive interactions belong to the same path only if they occur within a time window $\delta$, i.e., $t_2 - t_1 \leq \delta$.
Using these conditions, we can derive a set of paths $P$ from any temporal network.

In summary, the network topology constrains the paths that are possible in real-world systems, such as transport or communication systems. 
However, empirical path data contain additional information on the start- and end-points of paths and the specific sequences in which nodes are traversed. 
Thus, the real paths by which nodes indirectly influence each other can differ significantly from what we would expect solely based on the network topology.

\subsection{Modelling higher-order patterns in path data}\label{sec:modelling_higher_order_patterns}
In the previous section, we showed that empirical paths capture information not contained in the network topology. 
Based on our arguments, one might assume that paths are always better for capturing the dynamics on a networked system compared to the topology alone.
However, the validity of this argument strongly depends on the number of paths that we have observed.

\begin{figure}
    \centering
    \begin{tikzpicture}\sffamily
        \tikzstyle{path}=[decorate,decoration={triangles, shape size=3.5pt ,segment length=2.5pt}, line width=0pt, fill=black]
		\coordinate (vd) at (.5,-1);
		\def\s{1.035}
		
		\coordinate (A) at ($(-1,0)+(vd)+\s*(0,1.1)$);
		\coordinate (B) at ($(-1,0)+(vd)+\s*(0,-1.1)$);
		\coordinate (C) at ($(.25,0)+(vd)+\s*(0,0)$);
		\coordinate (D) at ($(1.5,0)+(vd)+\s*(0,0)$);
		\coordinate (E) at ($(2.75,0)+(vd)+\s*(0,1.1)$);
		\coordinate (F) at ($(2.75,0)+(vd)+\s*(0,-1.1)$);
		
		\coordinate (h) at (3.5pt,0);
		\coordinate (v) at (0,3.5pt);
		
		\coordinate (h2) at (2pt,0);
		\coordinate (v2) at (0,2pt);
		
		\foreach \x/\y in {A/C,B/C,C/D,D/E,D/F}{
			\draw[line width=19pt, black!15] (\x) to[out=0, in=180] (\y);
		}

		\draw[path, colorSG] ($(A)+2*(v)+2*(h)$) to[out=-20, in=200] ($(C)+2*(v)+2*(h)$);
		\draw[path, colorSG] ($(C)+2*(v)$) -- ($(D)+2*(v)$);
		\draw[path, colorSG] ($(D)+2*(v)-2*(h)$) to[out=-20, in=200] ($(E)+2*(v)-2*(h)$);
		
		\draw[path, customG] ($(A)+0*(v)+0*(h)$) to[out=0, in=180] ($(C)+0*(v)+0*(h)$);
		
		\draw[path, customY] ($(B)-2*(v)+2*(h)$) to[out=20, in=160] ($(C)-2*(v)+2*(h)$);
		\draw[path, customY] ($(C)-2*(v)$) -- ($(D)-2*(v)$);
		\draw[path, customY] ($(D)-2*(v)-2*(h)$) to[out=20, in=160] ($(F)-2*(v)-2*(h)$);
		
		\foreach \Point in {A,B,C,D,E,F}{
			\draw [line width=1pt, rounded corners=2pt, fill=white] ($(\Point)+4.5*(h2)-4.5*(v2)$) rectangle node {\textbf{\Point}} ($(\Point)-4.5*(h2)+4.5*(v2)$);
		}
    \end{tikzpicture}
    \caption{Exemplary set of paths on a network topology. We observe three colour-coded paths from $A$ to $B$ (\textcolor{customG}{\faSquare}), from $A$ to $E$ (\textcolor{colorSG}{\faSquare}), and from $B$ to $F$ (\textcolor{customY}{\faSquare}). The underlying network topology is shown in grey (\textcolor{black!15}{\faSquare}).}
    \label{fig:toy_example}
\end{figure}
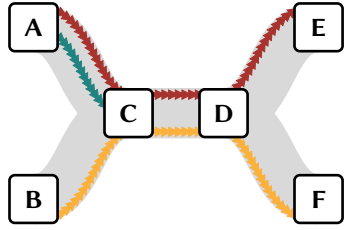

Let us consider the example shown in \Cref{fig:toy_example}.
As we can infer from the colour-coded paths, a path in $D$ will always continue to $E$ if it started in $A$.
In contrast, if the path started in $B$, it will continue to $F$.
But does this mean that paths from $A$ to $F$ do not exist, despite being possible according to the underlying network topology?
To address this question, we need to consider how often we observed the paths from $A$ to $E$ and $B$ to $F$.
If, e.g., we observed both paths only once each, we would have little evidence suggesting that a path from $A$ to $F$ is not be possible.
In a nutshell, we should not mistake absence of evidence as evidence of absence.
Hence, in this case, using the observed paths as indicators for all possible paths would overfit the data, and a network model that additionally accounts for theoretically possible but unobserved paths may be more appropriate.
In contrast, observing both paths many times without ever observing paths from $A$ to $F$ would indicate that paths from $A$ to $F$ do not exist or are at least significantly less likely than the observed paths. 
In this case, a network model would underfit the data by not adequately accounting for the patterns present in the empirical path data.

These examples underline that to capture the influence of nodes in real-world networked systems, neither a network model nor a limited set of observed paths is sufficient.
Instead, we require a model that can both capture the memory in the path data and allow transitions that are consistent with the network topology and cannot be ruled out because path data have not provided enough evidence.

\subsection{MOGen}\label{sec:MOGen}

Our work is based on \texttt{MOGen}, a multi-order generative model for paths~\citep{gote2020predicting} that combines information from multiple higher-order models.
In addition, \texttt{MOGen} explicitly considers the start- and end-points of paths using the special initial and terminal states $*$ and $\dagger$.
\texttt{MOGen} represents a path $v_1 \rightarrow v_2 \rightarrow \dots \rightarrow v_{l}$ as
\begin{align}
* \rightarrow v_1 \rightarrow (v_1, v_2) \rightarrow \dots\rightarrow (v_{l-K+1}, \ldots, v_{l}) \rightarrow \dagger,
\label{eq:path_representation}
\end{align}
where $K$ denotes the maximum memory the model accounts for.
Going back to the example from \Cref{fig:toy_example}, a MOGen model with $K=2$ would represent the \textcolor{colorSG}{\faSquare} path from $A$ to $E$ as
\begin{align}
* \rightarrow A \rightarrow (A, C) \rightarrow (A, C, D) \rightarrow (C, D, E) \rightarrow \dagger.
\label{eq:MOGen_example}
\end{align}
Hence, it would model the transition from $D$ to $E$ while considering if the path originated from $A$ or from $B$ (cf. our discussion in \Cref{sec:modelling_higher_order_patterns}).

Combining the representations of all paths in a set $P$, the resulting \texttt{MOGen} model is fully described by a multi-order transition matrix $\textbf{T}^{(K)}$ shown in \Cref{fig:MOGen_representation}.
The entries $\textbf{T}^{(K)}_{ij}$ of $\textbf{T}^{(K)}$ capture the probability of a transition between two higher-order nodes.

Considering no memory, a \texttt{MOGen} model with $K=1$ is equivalent to a network model but for nodes $*$ and $\dagger$ that additionally consider the starts and ends of paths.
In turn, a \texttt{MOGen} model with $K$ matching the maximum path length observed in $P$ is a lossless representation of the set of paths.
Thus, \texttt{MOGen} allows us to find a balance between the network model---allowing all observed transitions in any order---and the observed set of paths---only allowing for transitions in the order in which they were observed.

\subsubsection{MOGen: Fundamental matrix}

\input{MOGen_representation}

Building on the original model \citep{gote2020predicting}, we interpret the multi-order transition matrix $\textbf{T}^{(K)}$ of \texttt{MOGen} as an absorbing Markov chain where the states $(v_1, \ldots, v_{n-1}, v_n)$ represent a path in node $v_n$ having previously traversed nodes $v_1, \ldots, v_{n-1}$.
Using this interpretation allows us to split $\textbf{T}^{(K)}$ into a transient part $\textbf{Q}$ representing the transitions to different nodes on the paths and an absorbing part $\textbf{R}$ describing the transitions to the end state $\dagger$.
We can further extract the starting distribution $\textbf{S}$.
All properties are represented in \Cref{fig:MOGen_representation}.

This representation allows us to compute the Fundamental matrix $\textbf{F}$ of the corresponding Markov chain.
\begin{align}
	\textbf{F} = \left(\textbf{I}^{(m\times m)} - \textbf{Q}\right)^{-1}
\end{align}
Here, $\textbf{I}^{(m\times m)}$ is the $m \times m$ identity matrix, where $m$ is the number of nodes in the multi-order model without counting the special states $*$ and $\dagger$.
Entries $(i, j)$ of this Fundamental matrix $\textbf{F}$ represent the expected number of times a path in node $i$ will visit node $j$ before ending.
The Fundamental matrix $\textbf{F}$ is essential as it allows us to compute path centrality measures for the \texttt{MOGen} model \emph{analytically}.

\subsection{Centrality measures}\label{sec:centrality_measures}

We now introduce five \texttt{MOGen}-based centrality measures that we use throughout the subsequent analyses.
For all \texttt{MOGen}-based centrality measures, we also introduce the corresponding measures for a network and a path model.

\subsubsection{Betweenness centrality}
Betweenness centrality considers nodes as highly influential if they frequently occur on paths connecting pairs of other nodes.
\paragraph{Network model}
In a network, the betweenness centrality of a node $v$ is given by the ratio of shortest paths $\sigma_{st}(v)$ from $s$ to $t$ through $v$ to all shortest paths from $s$ to $t$ $\sigma_{st}$ for all pairs of nodes $s$ and $t$:
\begin{align}
    b_v = \sum_{s,t \in V}\frac{\sigma_{st}(v)}{\sigma_{st}}.
\end{align}

\paragraph{Path model}
Standard betweenness centrality calculated in a network model relies on the assumption that only the shortest paths are used to connect two nodes.
Using path data, we can drop this assumption and consider paths that are \emph{actually} used.
Therefore, we can obtain the betweenness of a node in a given set of paths $P$ by simply counting how many times a node appears between the first and last node of all paths.

\paragraph{MOGen}
For \texttt{MOGen}, we can utilise the properties of the Fundamental matrix $\textbf{F}$.
Entries $(v,w)$ of $\textbf{F}$ represent the number of times we expect to observe a node $w$ on a path continuing from $v$ before the path ends.
Hence, by multiplying $\textbf{F}$ with the starting distribution $\textbf{S}$, we obtain a vector containing the expected number of visits to a node on any path.
To match the notions of betweenness for networks and paths, we subtract the start and end probabilities of all nodes, yielding
\begin{align}
    b_v = \left(\textbf{S}\cdot\textbf{F}\right)_v - \textbf{S}_v - \textbf{R}_v.\label{eq:betweenness}
\end{align}
\Cref{eq:betweenness} allows us to compute the betweenness centrality for all nodes in the \texttt{MOGen} model---i.e., higher-order nodes.
The betweenness centrality of a first-order node $v$ can be obtained as the sum of the higher-order nodes ending in $v$.

\subsubsection{Closeness centrality (harmonic)}
When considering the closeness centrality of a node $v$, we aim to capture how easily node $v$ can be reached by other nodes in the network.

\paragraph{Network model}
For networks, we are therefore interested in a function of the distance of all nodes to the target node $v$.
The distance matrix $\textbf{D}$ capturing the shortest distances between all pairs of nodes can be obtained, e.g., by taking powers of the binary adjacency matrix of the network where the entries at the power $l$ represent the existence of at least one path of exactly length $l$ between two nodes.
This computation can be significantly sped up by using graph search algorithms such as the Floyd-Warshall algorithm \citep{floyd1962algorithm} used in our implementation.
As our networks are based on path data, the resulting network topologies are directed and not necessarily connected.
We, therefore, adopt the definition of closeness centrality for unconnected graphs, also referred to as harmonic centrality \citep{marchiori2000harmony}.
This allows us to compute the closeness centrality of a node $v$ as
\begin{align}
    c_v = \sum_{i\in V, i\neq v}\frac{1}{\textbf{D}_{vi}},\label{eq:closeness_centrality}
\end{align}
where $\textbf{D}_{vi}$ is the entry in the $v$-th row and $i$-th column of $\textbf{D}$.

\paragraph{Path model}
For paths, the distance between two nodes $v$ and $w$ can be obtained from the length of the shortest sub-path starting in $v$ and ending in $w$ among all given paths, also resulting in a distance matrix $\textbf{D}$.
Again, the closeness centrality is then computed using \Cref{eq:closeness_centrality}, however, with the modified distance matrix.

\paragraph{MOGen}
As \texttt{MOGen} models contain different higher-order nodes, $\textbf{D}$ captures the distances between higher-order nodes based on the multi-order network topology considering temporal correlations up to length $K$.
While we aim to maintain the network constraints set by the multi-order topology, we are interested in computing the closeness centralities for first-order nodes.
We can achieve this by projecting the distance matrix to its first-order form, containing the distances between any pair of first-order nodes but constrained by the multi-order topology.
For example, for the distances $d\{(A, B), (C, A)\} = 3$ and $d\{(B, B), (C, A)\} = 2$, the distance between the first-order nodes $B$ and $A$ is $2$.
Hence, while for the network, the distances are computed based on the shortest path assumption, multi-order models with increasing maximum order $K$ allow us to capture the tendency of actual paths to deviate from these shortest paths.
Based on the resulting distance matrix $\textbf{D}$, the closeness centrality is again computed following \Cref{eq:closeness_centrality}.
Therefore, while for all representations, we compute the closeness centrality of a node using the same formula, the differences in the results originate from the constraints in the topologies considered when obtaining the distance matrix $\textbf{D}$.

\subsubsection{Path end}
The path end $e_v$ of a node $v$ describes the probability of a path to end in node $v$.

\paragraph{Network model}
Path end cannot be computed for a network model as the information on the start and end of paths is not captured by this representation.

\paragraph{Path model}
For paths, $e_v$ is computed by counting the fraction of paths ending in node $v$.

\paragraph{MOGen}
For \texttt{MOGen}, all paths end with the state $\dagger$.
Therefore, $e_v$ is obtained from the transition probabilities to $\dagger$ of a single path starting in $*$.
This last transition can---and is likely to---be made from a higher-order node.
We can obtain the path end probability for a first-order node by summing the path end probabilities of all corresponding higher-order nodes.

\subsubsection{Path continuation}
When following the transitions on a path, at each point, the path can either continue or end.
With the path continuation $f_v$, we capture the likelihood of the path to continue from node $v$.

\paragraph{Network model}
As path information is required, no comparable measure exists for networks.

\paragraph{Path model}
Similarly to the path end, we obtain path continuation from a set of paths $P$ by counting the fraction of times $v$ does not appear as the last node on a path compared to all occurrences of $v$.

\paragraph{MOGen}
For \texttt{MOGen}, path continuation is given directly by summing the probabilities of all transitions in the row of \textbf{$\textbf{T}^{(K)}$} corresponding to node $v$ leading to the terminal state $\dagger$.
As for other measures, for \texttt{MOGen}, the continuation probabilities are computed for higher-order nodes.
We can obtain continuation probabilities for a first-order node $v$ as the weighted average of the continuation probabilities of the corresponding higher-order nodes, where weights are assigned based on the relative visitation probabilities of the higher-order nodes.

\subsubsection{Path reach}
Finally, we consider path reach.
With path reach, we capture how many more transitions we expect to observe on a path currently in node $v$ before it ends.

\paragraph{Network model}
Again, the path reach requires information on path ends.
Therefore, it cannot be computed using the network model.

\paragraph{Path model}
To compute the path reach $\rho_v$ for a node $v$, we average the number of remaining transitions before the path ends for all occurrences of $v$.

\paragraph{MOGen}
For \texttt{MOGen}, we can again use the properties of the Fundamental matrix $\textbf{F}$ and obtain the expected number of remaining transitions for any node $v$ as the row sum
\begin{align}
    \rho_v = -1 + \sum_{i\in V}\textbf{F}_{vi},
\end{align}
where $\textbf{F}_{vi}$ is the entry in the $v$-th row and $i$-th column of the Fundamental matrix $\textbf{F}$.
We subtract 1 to discount for the occurrence of node $v$ at the start of the remaining path.
Analogous to path continuation, we obtain the path reach of a first-order node $v$ by weighting the path reach of all corresponding higher-order nodes according to their respective relative visitation probabilities.

\section{Evaluating MOGen-based centralities in empirical path data}

In \Cref{sec:methods}, we argued that network models are likely to \textit{underfit} patterns in observed paths that are due to some paths occurring less often (or not at all), while others appear more often than we would expect based on the network topology alone.
Similarly, we expect the centralities computed directly on the paths to \textit{overfit} these patterns.
We, therefore, expect that when computing centralities based on the network or the paths directly, we misidentify the nodes that are actually influential.
We further conjecture that the errors caused by overfitting are particularly severe if the number of observed paths is low, i.e., if we have insufficient data to capture the real indirect influences present in the complex system.

\begin{figure}
    \centering
    \includegraphics[width=\linewidth]{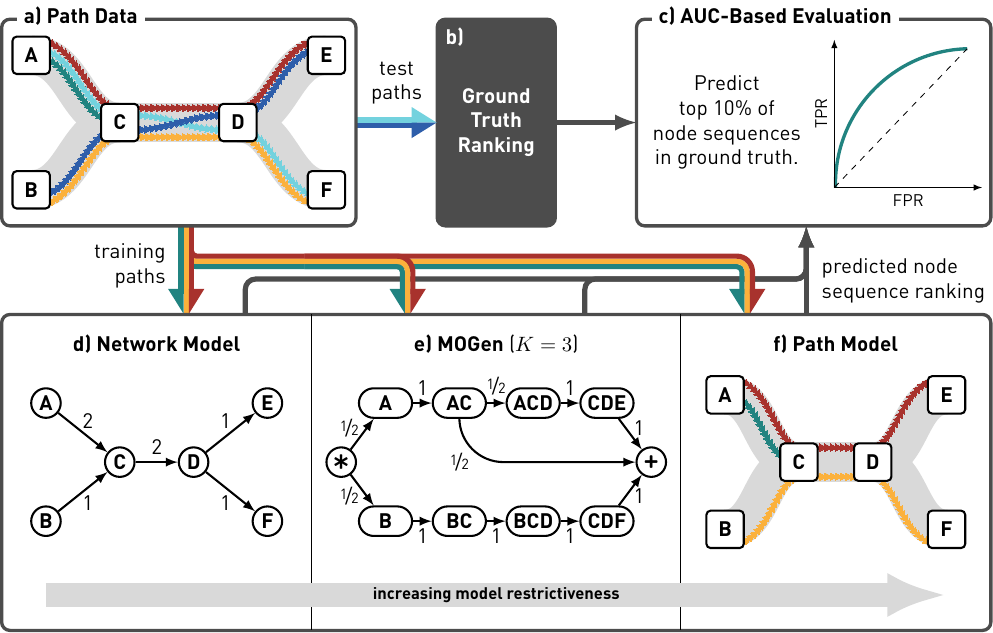}
    \caption{Overview of our approach to predict influential nodes and node sequences based on path data. We start from path data which we split into training and test sets. We learn three different models based on the training data: (i) a network model containing all transitions from the training data, (ii) a multi-order generative model containing observed higher-order transitions up to a maximum order of $K$, and (iii) a path model containing the full paths in the training set. Based on these models, we predict the influence of nodes or node sequences according to a broad range of centrality measures. We compare the ranking of node sequences to the ground truth rankings obtained from the test paths using AUC-based evaluation.}
    \label{fig:infographic}
\end{figure}

\subsection{Experimental setup}

We now test our \texttt{MOGen}-based centrality against network- and path-based measures in five empirical path datasets.
We refer to Appendix \ref{sec:appendix_path_datasets} for further information and summary statistics of these datasets.
For each path dataset, we compare three types of models:
First, a network model containing all nodes and edges observed in the set of paths.
Second, a path model which precisely captures the observed paths, i.e., the model is identical to the set of paths.
Third, \texttt{MOGen} models with different maximum orders $K$ that capture all higher-order patterns up to a distance of $K$.

We operationalise our comparison in a prediction experiment in which we aim to predict influential nodes and higher-order patterns in a set of test data based on training data.
\Cref{fig:infographic} provides an overview of our evaluation approach.

\subsubsection{Train-test split}
For our prediction experiment, we first split a given set of $N$ paths into a training and test set while treating all observed paths as independent.
We denote the relative sizes of the training and test sets as $\nicefrac{n_{\text{tr}}}{N}$ and $\nicefrac{n_{\text{te}}}{N}$, respectively.

\subsubsection{Ground truth ranking}
As introduced in \Cref{sec:methods}, our path-based centrality measures exclusively capture the influence of nodes in a set of observed paths.
While we expect this to lead to overfitting when making predictions based on training data, they yield precise ground truth influences when applied to the test data directly.
To obtain a ground truth ranking (see \Cref{fig:infographic}b), we sort the nodes and node sequences according to their influence in descending order.

\subsubsection{Prediction of influential nodes and node sequences} 

The network model is the least restrictive model for a set of paths.
In contrast, the path model always considers the entire history.
With $K=1$, a \texttt{MOGen} model resembles a network model with added states capturing the start- and end-points of paths.
By setting $K=l_{max}$, where $l_{max}$ is the maximum path length in a given set of paths, we obtain a lossless representation of the path data. 
By varying $K$ between $1$ and $l_{max}$, we can adjust the \texttt{MOGen} model's restrictiveness between the levels of the network and the path model.
We hypothesise that network and path models under- and overfit the higher-order patterns in the data, respectively, leading them to misidentify influential nodes and node sequences in out-of-sample data.
Consequently, by computing node centralities based on the \texttt{MOGen} model, we can reduce this error.

To test this, we train a network model, a path model, and \texttt{MOGen} models with $1 \leq K \leq 5$ to our set of training paths.
We then apply the centrality measures introduced in \Cref{sec:centrality_measures} to compute a ranking of nodes and node sequences according to each of the models.
In a final step, we compare the computed rankings to the ground truth ranking that we computed for our test paths.

\subsubsection{Comparison to ground truth}

While our models are all based on the same set of training paths, they make predictions for node sequences up to different lengths.
We allow the comparison of the different models' predictions through an upwards projection of lower-order nodes to their matching node sequences.
To this end, we match the prediction of the closest matching lower-order node $v_l \in \mathcal{L}$ as the prediction of the higher-order node $v_h \in \mathcal{H}$.
Here, $\mathcal{L}$ is the set of lower-order nodes, e.g., from the network model, whereas $\mathcal{H}$ is the set of higher-order nodes from the ground truth.
We define the closest matching lower-order node $v_l$ as the node with the highest order in $\mathcal{L}$ such that $v_l$ is a suffix of $v_h$.

We evaluate how well the predictions match the ground truth using an AUC-based evaluation approach.
Our approach is built on a scenario in which we aim to predict the top $10\%$ most influential nodes and node sequences in the ground truth data.
By considering this scenario, we transform the comparison of rankings into a binary classification problem, where for each node or node sequence, we predict if it belongs to the top $10\%$ of the ground truth or not.
All results reported throughout this manuscript refer to averages over at least five validation experiments.

\subsection{Comparison of the prediction quality}
\label{sec:results}

\begin{figure*}[t!]
    \centering
    \includegraphics[width=\textwidth]{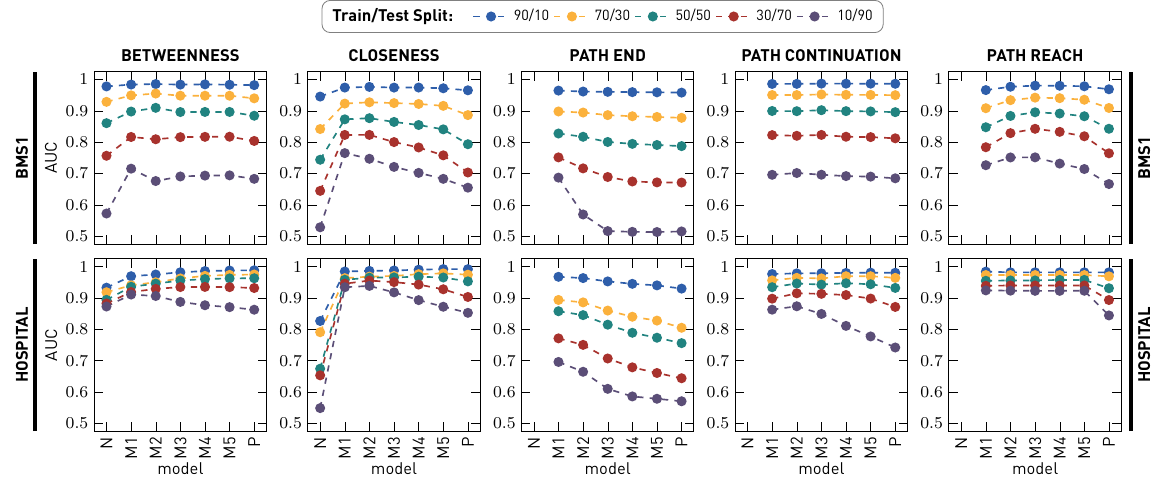}
    \caption{Prediction results for five centrality measures for the BMS1 and SCHOOL datasets and different train/test splits. N and P indicate the network and path model, respectively. M1 through M5 are \texttt{MOGen} models with maximum orders between 1 and 5.}
    \label{fig:results_network_measures}
\end{figure*}

We now present the results of our prediction experiments comparing the performance of network, path, and \texttt{MOGen} models to predict the influence of nodes and node sequences in out-of-sample data.
For ease of discussion, we start our analysis by focusing on the two datasets, BMS1 and HOSPITAL.
\Cref{fig:results_network_measures} shows the results for our five centrality measures.
For betweenness and closeness, we do not require information on the start- and end-point of paths.
Therefore, equivalent measures for the network model exist.
In contrast, no equivalent measures for the network model can be computed for path end, path continuation, and path reach.

We show the AUC values for the different models and for different relative sizes for our training and test sets.
The models shown on the $x$-axis are sorted according to the maximum distance at which they can capture indirect influences.
Thus, starting from the network model (N), via the \texttt{MOGen} models (M$K$) with increasing $K$, the models become more restrictive until ending with the path model (P).

Overall, the \texttt{MOGen} models outperform both the network model and the path models.
With less training data, the AUC scores of all models decrease.
However, as expected, these decreases are larger for the network and path models.
For the betweenness and closeness measures, this results in AUC curves that resemble ``inverted U-shapes''.
For the remaining measures, for which no equivalent network measures are available, we generally find that \texttt{MOGen} models with $K$ between 1 and 3 perform best, and the prediction performance decreases for more restrictive models, such as the path model.
Our results highlight the risk of underfitting for network models and overfitting for path models. 
We further show that this risk increases when less training data is available.

\begin{table*}[t!]
    \caption{AUC values for all models and measures on five datasets for a 30/70 train-test split. N and P indicate the network and path model, respectively. M1 through M8 are \texttt{MOGen} models with maximum orders between 1 and 8 (shown in \textcolor{customG!20}{\faSquare}). The best-performing result for each dataset and measure is highlighted in bold.}
    \label{tab:full_result_table}
    \centering
    \includegraphics[width=\textwidth]{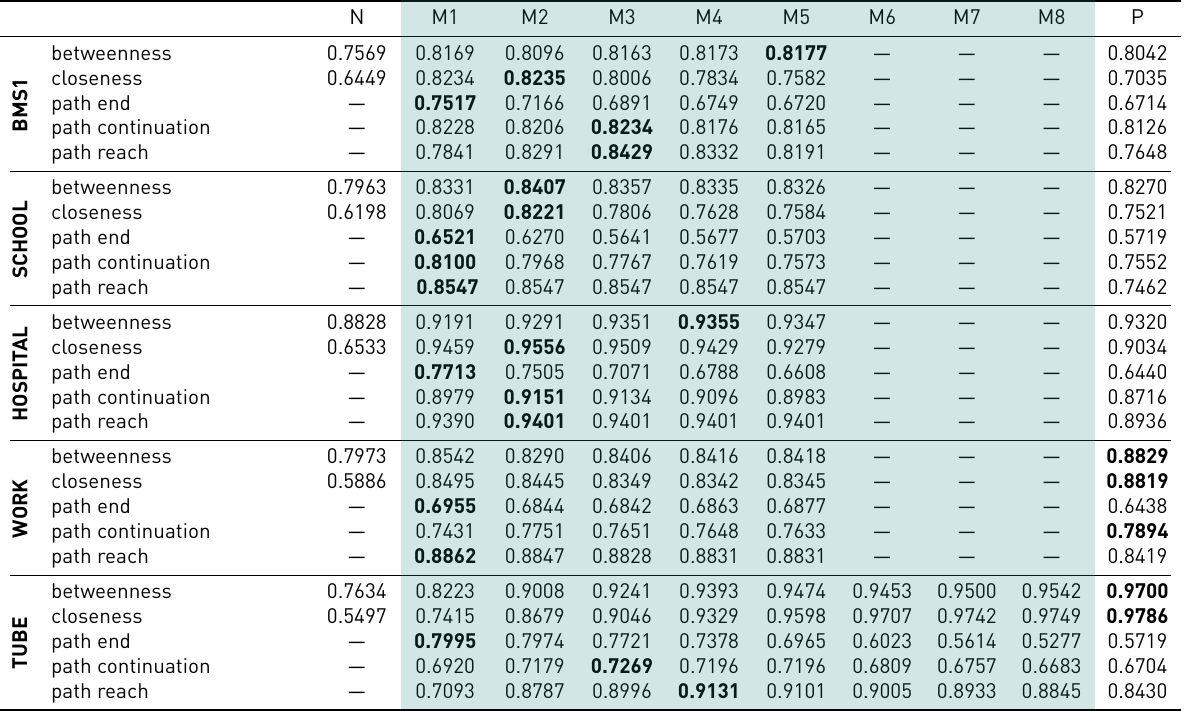}
\end{table*}

In \Cref{tab:full_result_table}, we show the results for all datasets and centrality measures for a 30/70 train/test split.
In general, we find similar patterns to those discussed with \Cref{fig:results_network_measures}.
However, for WORK and TUBE, the difference in prediction quality between the \texttt{MOGen} and path models decreases, and for some measures, the path model even yields better performance.
WORK and TUBE are those datasets for which we have the highest fraction of total observed paths compared to the number of unique paths in the datasets.
As shown in \Cref{tab:dataset_statistics}, BMS1 contains 59,601 total paths, of which 18,473 are unique.
This means that, on average, each unique path is observed 3.2 times.
These counts increase to 4 for SCHOOL, 4.6 for HOSPITAL, 6.7 for WORK, and 132.9 for TUBE.
The good performance of the path model for these datasets shows that the error we found with fewer observations is indeed due to overfitting.
In other words, if we have a sufficient number of observations, we can compute the centralities on the path data directly.
However, if the number of observations is insufficient, the path model overfits the patterns in the training data and consequently performs worse on out-of-sample data.
How many observations are required to justify using the path model depends on the number of unique paths contained in the dataset.

In conclusion, our results support our hypothesis.
By not capturing the higher-order patterns present in path data and not considering the start- and end-points of paths, the network model consistently underfits the patterns present in path data.
Similarly, the path model overfits these patterns.
Consequently, when using either model to rank the influence of nodes and node sequences in path data, we obtain rankings that are not consistent with out-of-sample observations.
Prediction performance can be significantly improved by using \texttt{MOGen} models that prevent underfitting by capturing higher-order patterns up to a distance of $K$ while simultaneously preventing overfitting by ignoring patterns at larger distances.

\section{Detecting community smells at genua}\label{sec:genua_analysis}

The data from \textit{genua} contains up to $\sim$2,000 paths for any one-year interval (cf. \Cref{fig:BugzillaRedmine_Aegis_Number_Of_Paths_comb}).
This means that analysing the path data directly is likely to \textit{overfit} the interaction patterns.
Therefore, in the following, we apply the \texttt{MOGen}-based centrality measures to identify community smells within the development process.
We visualise our approach in \Cref{fig:infographic2}.
In addition, we provide a detailed sequence of steps in \Cref{alg:community_smells_analysis}.

\begin{figure}[t!]
    \centering
    \includegraphics[width=\linewidth]{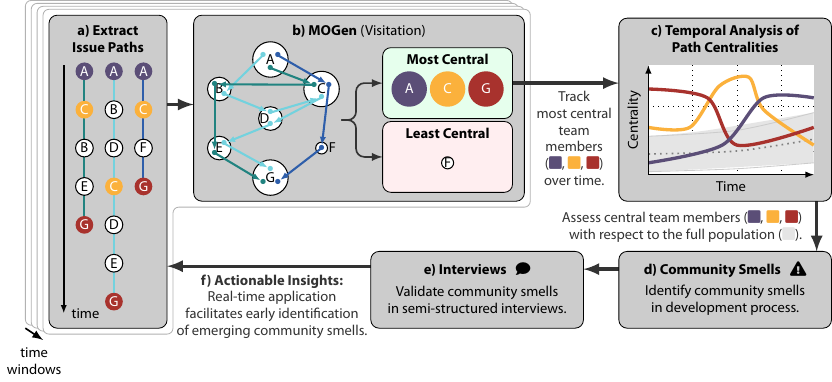}
    \caption{Overview of our approach to detect community smells in the development process at the German IT security company \textit{genua GmbH}. (a) For each one-year time window, we extract paths for all issues and their related changes. (b) We fit the \texttt{MOGen} model to the extracted paths and identify the most central team members according to the centrality measures introduced in \Cref{sec:centrality_measures}. (c) We identify those team members that are most central to the team and track them over time. (d) We identify community smells by comparing the centralities of those members with the values obtained for the remaining team. (e) We validate the detected community smells in semi-structured interviews with team members from \textit{genua}. (f) As our centrality measures are computationally effective, they can be employed in real-time to provide actionable insights on existing and emerging community smells to software development teams.}
    \label{fig:infographic2}
\end{figure}

\begin{table}[t!]\centering
\caption{Sequence of steps to detect community smells in the development process at the German IT security company \textit{genua GmbH}.}
\label{alg:community_smells_analysis}
\begin{tcolorbox}[width=\linewidth, colback=white, boxrule=1pt, boxsep=10pt, left=0pt, right=0pt]
\begin{algorithmic}\footnotesize\sffamily
\State \phantom{1}1. \textit{Start with development platforms storing team's time-stamped actions, labeled by issue.}
\State \phantom{1}2. \textit{Extract all actions from the databases.}
\State \phantom{1}3. \textit{Sort all actions according to their time-stamp.}
\State \phantom{1}4. \textit{Aggregate them at issue-level to form sequence of actions, i.e., paths.}
\State \phantom{1}5. \textit{Assign paths to rolling time-windows (one year long with three month shifts, cf. \Cref{sec:data}).}
\State \phantom{1}6. \textit{Fit a MOGen model for each time-window.}
\State \phantom{1}7. \textit{Compute path centralities for team members (cf. \Cref{sec:centrality_measures}).}
\State \phantom{1}8. \textit{Identify members consistently deviating from average centrality (cf. \Cref{sec:identifying_central_developers}).}
\State \phantom{1}9. \textit{Examine these members' centrality score time-series.}
\State 10. \textit{Detect time-series anomalies.}
\State 11. \textit{Explain anomalies, forming community smell hypotheses.}
\State 12. \textit{Validate hypotheses through semi-structured interviews.}
\end{algorithmic}
\end{tcolorbox}
\end{table}

\subsection{Higher-order interaction patterns}\label{sec:higher_order_patterns}

First, we infer the optimal maximum order for the issue tracker and code review data using \texttt{MOGen}'s built-in model selection approach.
To account for changes over time, we fit a separate \texttt{MOGen} model to all paths starting in a given one-year period.
We then move the one-year window by three-month increments and repeat the process.

For the issue tracker, we detect an optimal maximum order of one over the entire data.
We conclude that issue tracker interactions occur relatively unrestricted.
This is aligned with our knowledge of the development process at \textit{genua}.
Importantly, even a first-order \texttt{MOGen} accounts for the start- and end-points of paths.
The results presented later will show that the consideration of these start- and end-points, neglected by static network models, are essential.

For the code review process, we consistently find second-order patterns over the entire 15-year observation period\footnote{The code review data only starts approximately five years after the data from the issue trackers.}.
This indicates, that the subsequent step in the code review process is influenced by the previous two steps.
This implies that to accurately predict who will perform the integration task---incorporating code changes into the main codebase---information about both the developer and the reviewer of a change is necessary.

Based on these results, we analyse the issue tracker and code review data using first-order and second-order \texttt{MOGen} models, respectively.

\subsection{Identifying community smells in software development processes}\label{sec:identifying_central_developers}

Our aim is to identify lone wolves, bottlenecks, organisational silos and code-red situations in \textit{genua}'s development process---i.e., individual developers or small subgroups taking over specific tasks that nobody else takes over.
To this end, we compute the five centrality measures introduced in \Cref{sec:centrality_measures}.
In addition to these five measures, we also compute \emph{visitation}, capturing the frequency of observing each team member in any position on a path.
As a result, we obtain individual time-series for all team members and all centrality measures over the 20 years.

To identify those team members that are more likely to create community smells, we focus on individuals exhibiting consistently high deviations from the team's average centrality scores.
This is because such deviations suggest these members may have unique or unbalanced roles within the team, potentially leading to communication or collaboration issues that contribute to the emergence of community smells.
To this end, we define a deviation score $S_{p,i}$, which aggregates over all time-points $t\in T$ and centrality measures $c\in \mathcal{C}$:
\begin{align}\label{eq:focus_selection}
    S_{p,i} = \sum_{t\in T}\sum_{c\in\mathcal{C}}\left\vert\frac{v_{p,c,i}(t) - \bar{v}_{p,c}(t)}{\bar{v}_{p,c}(t)}\right\vert
\end{align}
In \Cref{eq:focus_selection}, $i$ identifies the developer for whom the deviation is computed and $p \in P = \{IT, CR\}$ denotes the development platform, i.e., issue tracker or code review, for which the score is computed.
$v_{p,c,i}(t)$ and $\bar{v}_{p,c}(t)$ represent the centrality value of a developer and the mean centrality value across all developers at time $t$, respectively.
Aggregating across the two platforms, we get a final score $S_i$ for each developer as
\begin{align}\label{eq:focus_aggregation}
    S_i = \frac{1}{\vert P\vert}\sum_{p\in P} S_{p,i}.
\end{align}
Based on \Cref{eq:focus_aggregation}, we select five team members ($A$--$E$) with the highest values for $S_i$ as the focal point of our analysis.

\subsubsection{Community smells in the issue tracking process}\label{sec:knowledge_islands_issue_tracker}

\begin{figure}[b!]
    \centering
    \includegraphics[width=\textwidth]{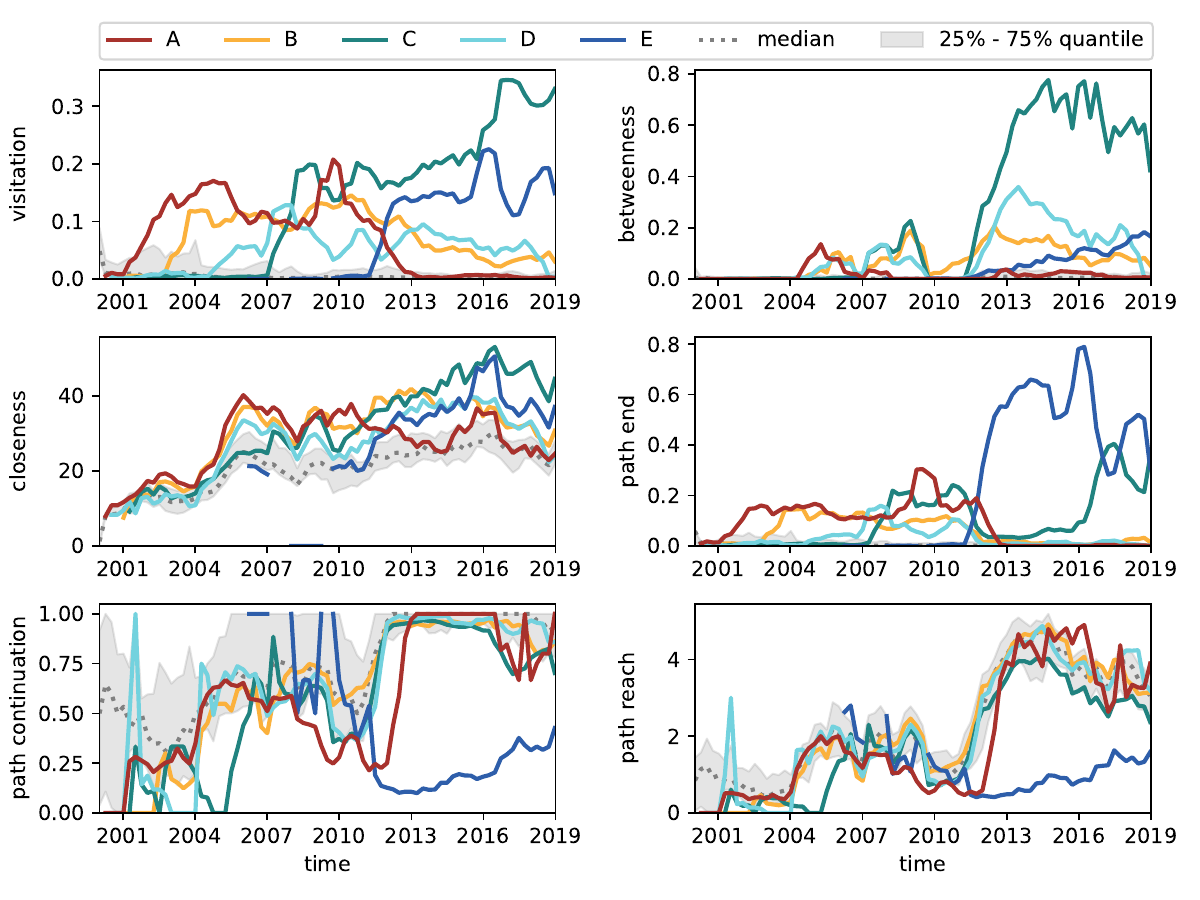}
    \caption{Centrality values for the five selected team members in the issue tracker data (\textit{Bugzilla}/\textit{Redmine}) over time.
    The significant increases of betweenness and average path reach in 2011 point to a notable change in the development process at that time.
    As indicated by path end, path continuation, and path reach, $E$ has held the final position in 60-80\% of all paths since 2011.
    The high visitation and betweenness of $C$ identify $C$ as a highly active and important transmitter of information within the team, although other team members also frequently appear in similar positions.}
    \label{fig:Centralities_BugzillaRedmine}
\end{figure}

\Cref{fig:Centralities_BugzillaRedmine} visualises the centrality values for the five selected team members in the issue tracking data.

\textbf{\textit{Visitation}.}~From the visitation, we observe that all selected team members have high activity levels.
$A$ and $B$ have high visitation probabilities until 2011, then are surpassed by $C$ and $E$.
Developer $C$ is responsible for over 30\% of actions from 2016 onward.
With the team having over 50 active team members during this period, this is a notable finding.
We also observe a transition in activity from $A$ to $E$, with $A$ dropping to 0\% and $E$ reaching around 15\% from 2012.

\textbf{\textit{Betweenness}.}~Betweenness centrality allows us to capture a team member's importance as a transmitter of information.
We observe a substantial change in betweenness centrality in 2011, with many team members' scores increasing significantly, especially for $C$ and $D$.
This change could be due to longer paths on the issue tracker, which would result in more team members appearing in the middle of a path.

\textbf{\textit{Closeness}.}~The above-average values for closeness centrality show that all highlighted team members are central to the development process.
It further indicates that they closely collaborate with a large part of the remaining team.
When team members change roles and become less active (e.g., $A$ in 2012), they deviate back towards the median.

\textbf{\textit{Path end.}}~Path end quantifies how frequently team members appear at the end of a path.
From \Cref{sec:genua_development_process}, we know that for the issue tracker, this relates to quality assurance and issue closing.
Path end reveals that starting in 2011, $E$ appears at the end of over 60\% of all paths, with other members' appearances significantly reduced.
This changed in 2016 when also $C$ appeared more frequently in this position.

\textbf{\textit{Path continuation and reach.}}~Finally, path continuation indicates if the development process continues after an action from a team member, and path reach quantifies how many steps are still left.
Here, we find that most highlighted developers show around average values.
However, as already indicated by path end, $E$ represents an exception to this as the development process rarely continues after $E$, and even if it does, it still only does so for a few steps.

\paragraph{Hypotheses for the issue tracking process}
We summarise our findings as a set of hypotheses regarding the issue tracking process.
Firstly, the significant increase of betweenness and average path reach in 2011 points to a notable change in the development process at that time.
As $E$ joined the team in 2011 and has held the final position in 60-80\% of all paths for over 10 years since we hypothesise that $E$ took on a new role in quality assurance or issue closing.
Therefore, we hypothesise that the knowledge of this aspect of the development process is primarily concentrated in $E$, potentially creating a bottleneck or organisational silo.

In addition, we have identified $C$ as a highly active and important transmitter of information within the team.
However, other team members also frequently appear in the same positions as $C$.
Consequently, based solely on the current data, we cannot definitively determine if $C$ contributes to a community smell.

\subsubsection{Community smells in the code review process}\label{sec:community_smells_code_review}

In \Cref{fig:Centralities_Aegis}, we show the centrality values for the code review data.
Unlike in the issue tracker data, $E$ does not appear in the code review data at all.
The absence of $E$ from the code review data provides further evidence for the unique position of $E$ we hypothesised based on the issue tracker data.

\begin{figure}
    \centering
    \includegraphics[width=\textwidth]{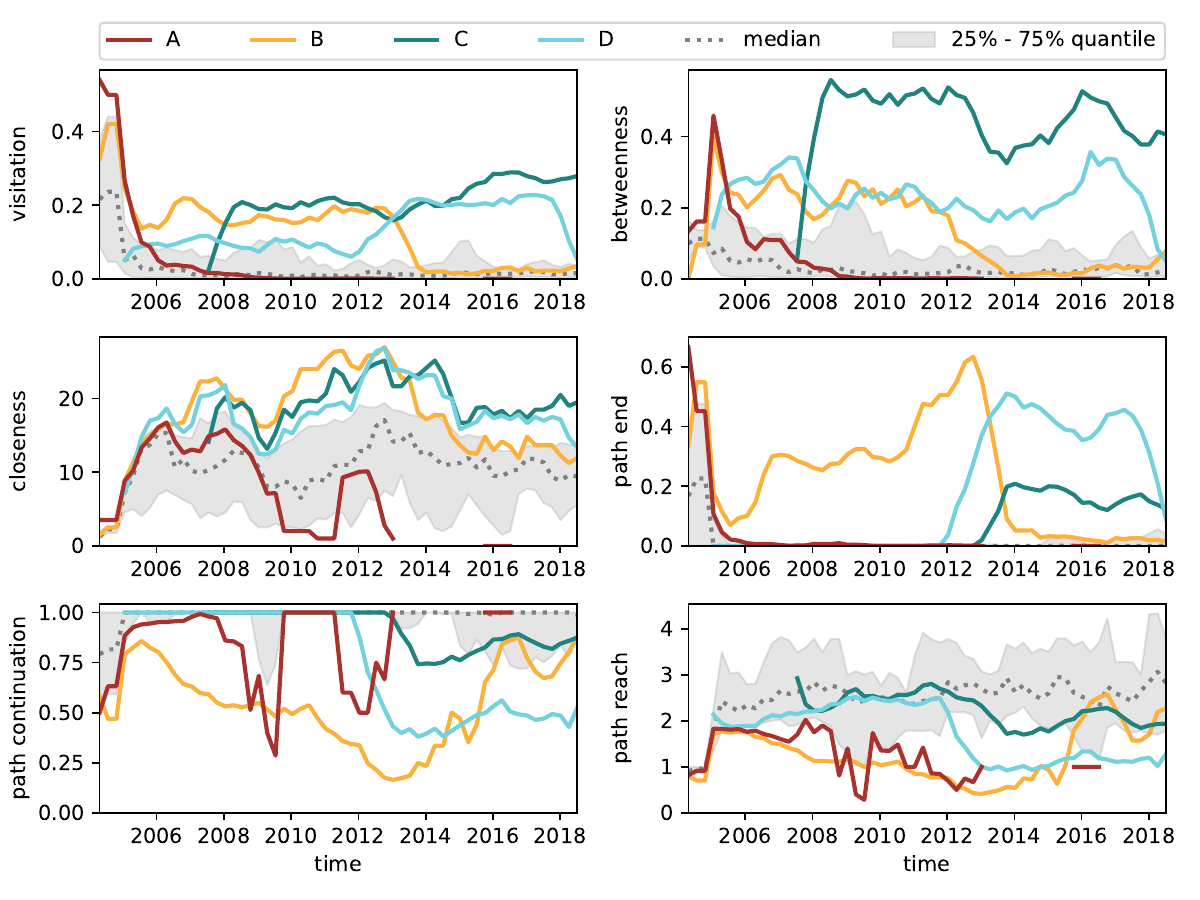}
    \caption{Centrality values for the five selected team members in the code review data (\textit{Aegis}) over time.
    Visitation shows that team members have similar maximum activity levels which they maintain over multiple years.
    Betweenness and path end can be interpreted as code review activity and integrations, respectively.
    Their time-series indicate that upon joining, team members initially focus on development and review tasks and take over integration tasks only after multiple years.
    From path end, we further observe that only very few people perform integrations at any point in time.}
    \label{fig:Centralities_Aegis}
\end{figure}

\textbf{\textit{Visitation}.}~Team members $A$--$E$ exhibit similar visitation probabilities and thus maximum activity levels, maintaining them over multiple years.
Their highest activity levels occur at different times, with clear handoffs between members (e.g., $B$ and $D$ from 2012 to 2014).
Similar to the issue tracker process, we again find $C$ to be very active and highly central.

\textbf{\textit{Betweenness and path end}.}~Based on our knowledge of \textit{genua}'s code review process (cf. \Cref{sec:genua_development_process}), we can interpret the betweenness centrality as code review activity and path end as integrations.
The betweenness shows that from 2007 onwards, $C$ performs more code reviews than any other team member.
Analysing the time-series of betweenness and path end together, we observe that upon joining the team, members initially focus on development and review tasks.
They only begin to take on integration tasks after several years with the team.
For instance, $C$ assumes this role after six years, while $D$ does so after seven years.
Notably, path end reveals that only $B$, $C$, and $D$ perform integration tasks (i.e., appear at the end of paths) at any given point in time.

\textbf{\textit{Closeness}.}~As for the issue tracker process, we find above-average closeness values of developers $A$--$E$.
Although we do not gain any additional insights regarding community smells at this stage, this substantiates the validity of our focus selection.

\textbf{\textit{Path continuation and reach}.}~Path continuation and path reach are below average for all highlighted team members.
This is consistent with their high path end, indicating that they perform integration tasks.

\paragraph{Hypotheses for the code review process}
We find that team members need multiple years of project experience to perform integration tasks.
This makes the team potentially vulnerable when experienced members leave.
With only very few people performing integrations at any point in time, we hypothesise that this represents a \emph{code-red situation}.

As for the issue tracker, we have identified $C$ as highly active and an important transmitter of information.
Even the study of node centralities in the code review process has not given us conclusive insights on the role of $C$, therefore motivating further analyses.
Uniquely, our \texttt{MOGen}-based centralities allow us to quantify the importance of not only nodes but also \emph{edges} in the network.
In \Cref{sec:appendix_second_order_centralities}, we examine the centralities of all edges where $C$ is either the target or the source to determine the extent of $C$'s interactions with various team members.
Through this analysis, we assess whether $C$'s interactions are limited to a small subset of the team or if they involve a broader range of team members.
This ultimately helps us understand the extent of knowledge sharing involving $C$.
Our results show that while $C$ is holding significant project knowledge, $C$'s broad interactions with various team members, both as sender and recipient, indicate that $C$ is not a lone wolf, and the team likely has active measures in place to promote knowledge exchange.
Therefore, we do \emph{not} consider $C$ as a potential community smell.

\section{Validating our results in semi-structured interviews}\label{sec:genua_interviews}

Based on the analysis of the data, we have hypothesised that: (i) $E$ constitutes a bottleneck or organisational silo in the issue tracker process related to either the quality assurance process or the closing of issues. In addition, we found that (ii) the integration task in the code review process represents a code-red situation, as only very few team members with extensive experience perform it.
However, based on the second-order interaction patterns, we conjectured that (iii) the team has measures in place to promote broad interactions between team members.

To test these hypotheses and simultaneously validate the performance of our evaluation approach, we conducted semi-structured interviews with five members of \textit{genua} that were or currently are members of the analysed team.
In particular, we had the opportunity to interview three of the highlighted team members---$B$, $D$, and $E$.
Besides those, we spoke with two additional team members, which we refer to as $F$ and $G$.
All interviews lasted approximately one hour and were conducted without the aid of any supplementary material.
The interviews were conducted in March of 2021 in German, with the transcripts subsequently being translated into English.
Following each interview, we debriefed the interviewee and discussed the set of figures shown throughout the previous sections.
We performed all evaluations and the selection of highlighted team members solely based on their pseudo-anonymised ID (cf. \Cref{sec:path_extraction_genua}).
After sending our results to \textit{genua}, the team de-anonymised the five highlighted team members for the subsequent interviews.

All interviews started with general questions about the team and the interviewee's role and career within the project.
Based on our quantitative findings, we designed our subsequent questions around the two community smells we detected.
Following this structure, we first report the responses to questions aimed at validating that our approach is capable of identifying those team members most central to the development process.
Subsequently, we present our findings concerning the community smell in the issue tracker process.
Finally, we summarise the responses to our questions addressing the code-red situation in the integration task during the code review process.

\subsection{Essential team members}

To validate our approach of detecting influential team members using centrality measures, we asked every interviewee to name three team members essential for the project, which cannot include themselves.
The aggregated results from all five interviews are shown in \Cref{tab:interview_importance}.

In total, only two interviewees mentioned a team member not highlighted in our data.
Interestingly, both interviewees mentioned the same team member, which we refer to as $H$.
When we asked why this team member was critical, we learned that $H$ took over a large part of the work from $D$ after $D$ changed teams in 2018.
As our data ends at the end of 2018, most of this period is outside our observation period.
Thus, overall, we could confirm that our approach is effective at detecting essential team members.

\begin{table}[t!]
    \caption{Number of times each team member was mentioned when asking our five interviewees to name the three team members most important for the development process (excluding themselves). The congruence with our highlighted team members is remarkable as, in total, 176 different people contributed to either the issue tracker or the code review platform during our 20-year observation period.}
    \label{tab:interview_importance}
    \centering
    \begin{tabularx}{.5\textwidth}{CCCCC|c}
        \toprule
        \textbf{A} & \textbf{B} & \textbf{C} & \textbf{D} & \textbf{E} & \textbf{other (H)} \\
        \midrule
        1 & 4 & 4 & 3 & 1 & 2 \\
        \bottomrule
    \end{tabularx}
\end{table}

The counts with which different team members were mentioned also match our expectations.
$A$ who stopped contributing in 2012 is mentioned only once.
Instead, most interviewees consider current or more recent team members as more important.
That said, $E$, whom we hypothesise to be a bottleneck or organisational silo, is only mentioned once.
We further discuss this result in the following section.

\subsection{The quality assurance process}

In \Cref{sec:knowledge_islands_issue_tracker}, we found that both the betweenness centrality and the path reach increased substantially for some of the team members.
We hypothesised that this is due to a newly introduced or significantly extended quality assurance process headed by $E$.
However, we could not fully rule out that the increases were related to the transition from \textit{Bugzilla} to \textit{Redmine}, which occurred around 2010.

As we found during our interviews, the transition from \textit{Bugzilla} to \textit{Redmine} did not modify the development process.
\begin{quote}
	``The process didn't change much. The categories and priorities may have looked minimally different, but the process itself [...] there nothing has changed.'' (statement by $D$)
\end{quote}
$B$ confirmed that the development process itself did not change, arguing that
\begin{quote}
	``[t]his [\textit{Bugzilla}] was just a tool.'' (statement by $B$)
\end{quote}
When asking for the reason behind the transition, we learned that
\begin{quote}
	``[t]he switch [from \textit{Bugzilla} to \textit{Redmine}] was made mainly because they wanted to track features that were promised to customers, which was not possible with \textit{Bugzilla}.'' (statement by $E$)
\end{quote}
Further, the team is not aware of any causal relationship between the transition and $E$ joining the team:
\begin{quote}
	``No, I think this was pure chance.'' (statement by $E$)
\end{quote}

Thus, we conclude that the transition from \textit{Bugzilla} to \textit{Redmine} was not accompanied by any significant change in the development process.

Next, we aimed to determine how the quality assurance process worked before and since $E$ joined the team.
Here, we found that before $E$ joined, the team used an internal testing setup:
\begin{quote}
	``A unique aspect of \textit{Aegis} is that for each bugfix or feature that you develop, you have to write a test. You then need to execute this test twice---once with your changes and once without your changes. Without your changes, the test needs to fall on its nose and fail. With your changes, the feature is now there, and the test is successful. Through this process, we have already tested the code. Not only developed but simultaneously also tested.'' (statement by $D$)
\end{quote}
In addition to the testing by the development team,
\begin{quote}
	``[t]he customer managers have a natural interest to test new features before installing them for a customer.'' (statement by $B$)
\end{quote}
However, with both the product and the customer base growing significantly, the team eventually introduced an external quality assurance process:
\begin{quote}
	``The BSI [Federal Office for Information Security] responsible for the certification of [the product] requested an additional quality assurance process conducted by an external person not part of the development team. I became this person.'' (statement by $E$)
\end{quote}

Over the last ten years, $E$ has developed a meticulous testing setup allowing the evaluation of new releases in environments similar to those used by \textit{genua}'s clients.
However, as we hypothesised,
\begin{quote}
	``[n]obody apart from $E$ knows how the testing environment works.'' (statement by $B$)
\end{quote}
$E$ confirmed this, stating:
\begin{quote}
	``I am effectively the entire quality assurance department for [the product]. In moments where I don't feel like doing it, go on holiday, or am busy with other things, it [new changes] remains without quality assurance.'' (statement by $E$)
\end{quote}

So what happens if $E$ can no longer perform the work?
\begin{quote}
	``Certainly, nobody would know my test environment and my tests at a deeper level. One or two people have performed tests of patches---this means they have effectively tested a new version of the software with the existing set of tests. However, I am the only one who knows the setup in depth and knows how to properly create new tests or adapt tests for new features.'' (statement by $E$)
\end{quote}
In other words, there would be severe consequences for the team:
\begin{quote}
 	``If [$E$] is absent or unable to perform the work, we have a massive problem. Quality assurance is undoubtedly something where we have [$E$] who has done this for many years and is genuinely the only one.'' (statement by $G$)
\end{quote}

Despite these consequences, only $G$, who has also worked in quality assurance, mentioned $E$ as one of the most important team members.
Also, no steps are taken to mitigate the potential consequences:
\begin{quote}
	``Concerning me, if there are any steps taken to moderate the consequences if I was no longer there? I don't know; I haven't witnessed any.'' (statement by $E$)
\end{quote}
We argue this is particularly critical as the external quality assurance process, which $E$ is responsible for, was explicitly requested by the Federal Office for Information Security and is thus required to obtain the product's certification.

In conclusion, the interviews confirm our hypothesis that $E$ is crucial for the quality assurance of the product.
We further learned that $E$ is the only team member with detailed knowledge of the testing environment.
Hence, our quantitative analysis correctly identified $E$ as a bottleneck in the issue tracker process and as an organisational silo regarding the quality assurance process.
Interestingly, when asked to name team members essential to the development process' functioning, $E$ was only named once.
This shows that especially the core team, responsible for the development of new features, is not actively aware of this community smell.
It also explains the lack of measures in place to mitigate the consequences emerging if, for any reason, $E$ drops out.

Overall, we validate that our method successfully detects community smells.
We were further able to show that, in this case, the team was not widely aware of the community smell and the corresponding risk.
But what can a team do to mitigate the risk?
To address this question, we next look at the integration task in the code review process.

\subsection{The integration of changes}

Our quantitative analysis revealed that the integration task is only performed by a few team members with multiple years of experience in the project.
However, our analysis of second-order nodes showed that these team members not only interact with each other but instead with a broad spectrum of other team members.
Therefore, we hypothesised that there are active processes present in the team to (i) ensure that there are sufficient team members to perform integrations at all times and (ii) facilitate the spread of knowledge among team members.

During our interviews, everyone confirmed that, indeed, to perform integrations, deep insights into the structure and development history of the product are required, which are very hard and time-consuming to obtain:
\begin{quote}
	``[The product] is now significantly older than 20 years, and it still exists, [...], but there were multiple generations of developers that have worked on it. Those are now gone again, which means that there is quite a lot of knowledge---especially undocumented knowledge---concentrated in just a few team members.'' (statement by $D$)
\end{quote}
This rich history also contains numerous design decisions that might look wrong initially but were made for good reasons.
\begin{quote}
	``With certainty, things that look peculiar at first glance [today] had to be done exactly like this for reasons that date back six, eight, or ten years. The longer you are part of the team, the more corners of the product you get to know, allowing you to make changes relatively efficiently without running the risk of breaking everything in another place.'' (statement by $E$)
\end{quote}
This complexity makes it very difficult for newcomers to get started contributing to the project, and even experienced team members have a hard time understanding all aspects and use cases:
\begin{quote}
	``I think the initial barrier of entry is very high for [the product]. This means that while experienced Unix, Pearl, and C developers can collaborate on a small subarea, they lack the bigger picture of how the product works as a whole and how it is deployed with our clients.'' \\
	``I would personally argue that even a long-term developer of [the product] could not put the product in operation or configure the product for a client. Even those developers don't know how the different components interact and how they, therefore, need to be configured as a whole.'' (statements by $F$)
\end{quote}

To cope with this complexity, the team has defined multiple roles corresponding to different tasks in the development process.
\begin{quote}
	``There are three different roles: the role of the developer, the role of the reviewer, and the role of the integrator. All those who participate are, in any case, first developers. A few less are reviewers. For example, new trainees are initially not allowed to review.'' (statement by $D$)
\end{quote}
To become an integrator, team members indeed need multiple years of experience working on the product.
\begin{quote}
	``We only assign the integrator role to experienced developers who see the bigger picture.'' (statement by $B$)
\end{quote}
We note that commonly a single team member has multiple roles, with experienced team members taking over all three roles depending on the need.

Over the 20 years of development history included in our analysis, we found that three to four team members perform integrations at all points in time.
However, this count is not something the team consciously keeps track of but rather a result of the code review process in which the developer, reviewer, and integrator of a change cannot be the same team member.
\begin{quote}
	``It's purely a practical problem as if one [integrator] is on holiday and another one is sick, nobody can integrate. Therefore, you need a third one that can integrate.'' (statement by $D$)
\end{quote}
If there is a lack of integrators, the tasks are given to the next most experienced team member.
\begin{quote}
	``You have to keep the process going. When you realise, ok, we have one resource that integrates and five developers waiting for someone to integrate, then you say, ok, who is the second most experienced now? If it still doesn't work, you say, who is the third most experienced now? And then you give the integrator role to so many people until it works.'' (statement by $B$)
\end{quote}

However, the constraints of the review process do not explain why the team does not form subgroups that continuously review and integrate the changes developed by each other.
Here $B$, who has been with the team for the full observation period, told us:
\begin{quote}
	``In the beginning, everyone had their own area. There was one who did the WWW relay, and the other one had to do the IP relay. When we wanted to do five features in one release, then everyone was assigned a different feature.'' \\
	``Then we restructured the whole thing by introducing the Scrum process where now the team as a whole is responsible for all things that are done. Thus we tried to break the whole thing up a bit.'' (statements by $B$)
\end{quote}
Thus, by introducing Scrum, the team actively and consciously tries to spread knowledge about the product throughout the team.
\begin{quote}
	``One of the philosophies of Scrum is that everyone can do everything to address exactly the problems arising when the bus comes [referring to the truck factor, which is also known as the bus factor], or Google simply pays more. Thus we try to counteract exactly these problems in advance through XP [Extreme Programming] and pair programming [deliberate pairing of team members with different expertise].'' (statement by $D$)
\end{quote}
In addition, the review process also contributes to knowledge sharing within the team.
\begin{quote}
	``Given the complexity of [the product], we only have relatively few developers. Whenever something is changed, someone has to look at it [review and integrate it], which means that there is inevitably a mix of what people see.'' (statement by $G$)
\end{quote}

However, while, as we could show, these efforts lead to a strong mixing of interaction partners, there remain some areas in which team members specialise.
\begin{quote}
	``However, I think there are still comfort zones where people make initial changes and whom you let do it [make changes in a certain area of the codebase].'' (statement by $G$)
\end{quote}
Also, as $D$ experienced in another project, even three experienced team members can leave in short succession, causing trouble for the remaining team.
\begin{quote}
	``I had the plan to switch from [the analysed project] to [the project I'm in now] for two months to have a look at the other project. They have a different programming language, a different framework etc. We were in the process of scheduling when we would start when one of the colleagues from that project quit. Then we said, ok, let's do that right away as then I'll get some more information from him. Because he quit, we decided that I'd switch completely. Afterwards, it didn't take long, and two more core developers from the team quit, and that was pretty hard. A lot of knowledge left very quickly, and it often happens that I read some code that I don't know and that I don't understand right away. Then it takes me a lot longer. Also, the other remaining core developers need to look these things up because it's not their expertise. In that situation, I felt the bus factor very hard. There you notice that there was quite a bit of sand in the gears.'' (statement by $D$)
\end{quote}
Thus, based on our \texttt{MOGen}-based centrality measures, we correctly identified the integration process as a code-red situation.

In conclusion, our quantitative analysis revealed that the team succeeded at enabling at least three team members to perform integrations at any point throughout our 20-year observation period.
Integrations are the most challenging part of the code development and review process.
They require an in-depth overview of the project that, due to its long and complex history, even experienced newcomers need multiple years to obtain.
To facilitate the diffusion of knowledge, we have identified three key measures employed by the team: 
First, a strict code review process in which all three parts of introducing a new change to the product must be performed by different team members.
Second, Scrum, where the whole team is involved and responsible for new milestones in the project.
Third, the team employs Extreme Programming and pair programming, where team members with different areas of expertise are deliberately paired to jointly develop a change and share knowledge in the process.

As we found during our quantitative analysis, these measures allow the team to significantly reduce the reliance on any individual team member.
We argue that applying these already established measures to the quality assurance process would also reduce the risk concentration in $E$ that we observed there.
However, it would also come with increased costs as an additional team member would need to be involved in the quality assurance process.
The resulting additional communication overhead would further lead to a reduction in individual productivity \citep{gote2019analysing}.
Finally, as we saw with the last example, risks can never be entirely eliminated.
Also, the importance of cascade processes, where the departure of one team member causes other team members to leave, needs to be further explored \citep{burkholz2018correlations, callaway2000network}.
Thus, the best strategy that a team can take is to continuously evaluate the team's risk concentrations, allowing it to be aware of the risks and act accordingly.
Through our interviews, we could validate that the centrality measures around the \texttt{MOGen} model proposed in this manuscript are a powerful, versatile, and fine-granular way to capture and quantify who holds knowledge and performs certain aspects of the development process and how this knowledge is shared within the team.

\section{Threats to validity}\label{sec:threats-to-validity}

We now discuss the threats to validity for our empirical study identifying community smells, which we discussed in \Cref{sec:data,sec:genua_analysis,sec:genua_interviews}.
 
First, applying our approach and the centrality measures to software development data ranging over more than 20 years yields an immense amount of detail, as shown in the result figures throughout this manuscript.
As mentioned in \Cref{sec:identifying_central_developers}, this amount of detail is both a major strength and weakness of our analysis.
On the one hand, it allows us to gain deep insights into any aspect of the development process.
However, on the other hand, it makes selecting the focus of our analysis challenging.
We have opted to perform a detailed analysis for a subset of team members.
Here, we specifically looked for instances where team members show extreme values---both low and high---for our centrality measures.
Our results demonstrate that this approach works well to identify essential members of the team.
However, as we found with the quality assurance process, team members are not necessarily aware of all community smells within the team.
Thus, while the community smells we identified were confirmed in our interviews, our analysis approach cannot guarantee that these are the only ones present in the team.

The second potential threat to validity originates from the data used for our analyses.
Here, we assessed all available data from the issue trackers and the code review platform.
In our interviews, we further confirmed that the vast majority of development processes appear on these platforms.
However, with personal communication, email, or online chat, there are additional communication channels used for knowledge exchange that are not available for our analyses.
We further learned that processes such as Scrum meetings or pair programming that we identified to be essential in avoiding community smells are not recorded and, therefore, do not appear in our data.
Through our interviews, we were able to corroborate our quantitative findings made without this data.
Nevertheless, there are likely other team members in crucial roles, such as the chairs of the Scrum meetings, that are not captured by the present data.
In future work, we will extend our community smell detection method to include additional data sources, e.g., by also tracking developers' activity directly in the codebase \citep{gote2019git2net}.

Third, throughout our analysis, we make the critical assumption that not observing a team member performing a task means that this team member cannot perform the task.
Conversely, we assumed that if we only observe a specific subset of team members performing a task, these are the only team members who can take over the task.
Basing our evaluation on this assumption, our quantitative approach cannot be used in an unsupervised manner to detect community smells entirely independently.
Instead, our method aims to inform software development teams and their managers on where community smells are likely to exist.
Subsequently, similar to our interviews, the team itself can perform an informed further investigation and draw the appropriate consequences.
We argue that this is the best we can do, as due to privacy and intellectual property considerations, we do not have any data on the content of interactions---i.e., the comments written by team members and the developed source code---and, as mentioned above, we are missing all data on personal interaction channels.
While this presents a threat to the external validity of our approach, it does not affect the results presented here, as all findings could be validated in our semi-structured interviews.

Finally, our methodological approach to detect community smells is based on our knowledge about the development process applied at \textit{genua} that we discussed in \Cref{sec:genua_development_process}.
This means that it is not possible to directly carry over our interpretations of the different centrality measures to teams using a different development process.
However, at its core, our method is able to identify central nodes and structural bottlenecks in any path data set.
Therefore, while the development processes might differ, we are convinced that our method is applicable to other teams.
In fact, it should apply to any system for which we can observe path data.
Nevertheless, this needs to be tested in an extended study considering data from additional teams and systems in future work.

\section{Conclusion}
\label{sec:conclusion}

In this paper, we proposed a new approach to locate community smells in software development processes using higher-order network centralities.
Our work expands the current understanding of social network analysis (SNA) by addressing the limitations of static network models and demonstrating the benefits of higher-order network analysis in capturing the dynamics of software development teams.

Based on the higher-order model \texttt{MOGen}, we proposed measures to quantify the influence of both nodes and node sequences in path data according to five different notions of centrality.
Our centrality measures range from simple concepts like betweenness to complex measures such as path reach.
We demonstrate in a prediction experiment with five empirical datasets, that utilising our \texttt{MOGen}-based centrality measures results in improved predictions of influential nodes in time-series data compared to both \emph{network} and \emph{path} models.

Finally, we showed how our method could be applied to detect community smells, i.e., sources of unforeseen project cost connected to a `suboptimal' community structure, within a product team at the German IT security company \textit{genua GmbH}.
With our analysis, we identified a developer acting as a bottleneck and organisational silo in the quality assurance process and a code-red situation in the code review process.
The validation of our findings through semi-structured interviews with \textit{genua} developers confirmed the presence of the identified community smells.
The team was already aware of one community smell and was actively addressing it.
However, they were not aware of the second community smell, which our approach helped uncover.
This means that our analysis enables the team to take counter measures against it.
This highlights the potential of our methodology in aiding software teams to identify and address hidden community smells, ultimately improving the overall development process.

In conclusion, this paper advances SNA by introducing higher-order network centralities to effectively capture dynamics and indirect relationships.
This enhances the detection of patterns not captured by static network models---the most popular models for relational data.
Our findings reveal that higher-order network centralities can effectively identify community smells that would remain undetected using traditional SNA.
The approach proposed in this work can be applied in real-time to warn the teams of community smells before they cause harm, emphasising the real-world value that higher-order models can provide.

\section*{Archival and Reproducibility}
Sources for all data used in this manuscript are provided. A reproducibility package is available at \url{https://doi.org/10.5281/zenodo.7139438}.
A parallel implementation of the \texttt{MOGen} model is available at \url{https://github.com/pathpy/pathpy3}.

\section*{Acknowledgments}
We thank the five anonymous interviewees from \textit{genua} for the valuable insights provided during the interviews.
C.G., V.P., and I.S. acknowledge support by the Swiss National Science Foundation, grant 176938.


\input{arXiv.bbl}
\clearpage

\begin{appendices}

\section{General path datasets}\label{sec:appendix_path_datasets}

We test our hypothesis in five empirical path datasets containing observations from three different categories of systems: (i) user clickstreams on the Web (BMS1: \citealp{gazelle}), (ii) travel itineraries of passengers in a transportation network (TUBE: \citealp{LTdata}), and (iii) time-stamped data on social interactions (HOSPITAL: \citealp{Vanhems2013}; WORKPLACE: \citealp{Genois2015_FaceToFace}; SCHOOL: \citealp{stehle2011high}).
BMS1 and TUBE are directly collected in the form of paths.
For SCHOOL, HOSPITAL, and WORKPLACE, we extracted paths following \Cref{sec:paths_on_network_topologies}, using $\delta$ as 800s, 1,200s, and 3,600s, respectively.
The raw data for all datasets are freely available online (cf. references above).
We provide summary statistics for all datasets in \Cref{tab:dataset_statistics}.

\input{dataset_description.tex}

\section{Interaction broadness at genua}\label{sec:appendix_second_order_centralities}

$C$ holds a significant amount of knowledge on the project, which, if not adequately spread within the remaining team, would make $C$ a lone wolf and possibly cause issues with future development.
Therefore, in this appendix, we study the broadness of $C$'s interactions with the rest of the team.

For this, we recall that in \Cref{sec:higher_order_patterns}, we detected an optimal order of $K=2$ for the code review process.
In other words, we found that in the code review process, previous interactions have a statistically significant impact on subsequent interactions.
Our \texttt{MOGen}-based centralities uniquely allow us to study the importance of these interactions, as, next to the centrality of nodes, we can also compute the centrality of edges.

In \Cref{fig:Centralities_Aegis_2nd_Order_devC,fig:Centralities_Aegis_2nd_Order_devC_part2}, we show the centralities of all edges where $C$ is either the target or the source, respectively.
With 12 target edges and 13 source edges with similar values for all centralities, our results clearly show that $C$'s interactions are not limited to a small subset of the remaining team.
Instead, $C$ interacts with a broad range of other team members both as sender and recipient.
This means that whenever $C$ develops or reviews code for a change, no specific other team member reviews or integrates it.
Similarly, $C$ reviews and integrates code from a wide range of other team members.

In conclusion, we find that $C$ is a highly central and prolific member of the team.
However, $C$ is heavily involved in interactions with a large part of the remaining team.
In addition, many other team members also appear in positions where $C$ is central.
This suggests that $C$ is \textit{not} a lone wolf.

We further show that similarly broad interaction patterns can be found for all other highlighted team members (see \Cref{fig:Centralities_Aegis_2nd_Order_devA,fig:Centralities_Aegis_2nd_Order_devA_part2,fig:Centralities_Aegis_2nd_Order_devB,fig:Centralities_Aegis_2nd_Order_devB_part2,fig:Centralities_Aegis_2nd_Order_devD,fig:Centralities_Aegis_2nd_Order_devD_part2})\footnote{Team member $E$ does not appear in the code review data and is, hence, not shown.}.
Finding such broad interactions for all analysed team members suggests that the team has active measures in place to promote knowledge exchange within the team.

\begin{figure}
    \centering
    \includegraphics[width=\textwidth, trim=0 120 0 0, clip]{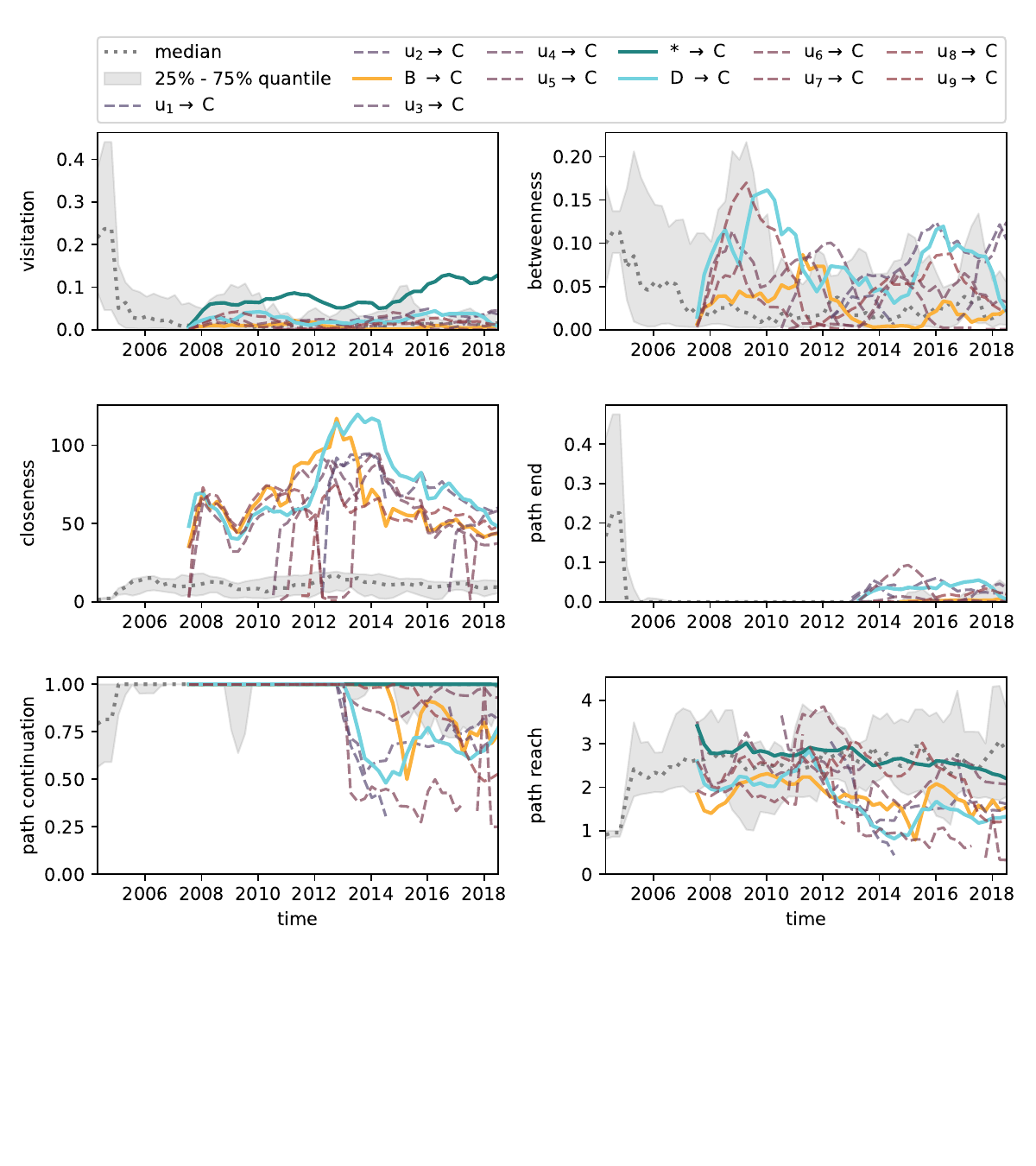}
    \caption{Centrality of all edges in the \texttt{MOGen} model with $K=2$ in the \textit{Aegis} data.
All edges to $C$ that reach a total visitation of at least 2\% in any given year are highlighted.
The median and the 25\% to 75\% quantile are shown in grey.}
    \label{fig:Centralities_Aegis_2nd_Order_devC}
\end{figure}

\begin{figure}
    \centering
    \includegraphics[width=\textwidth, trim=0 120 0 0, clip]{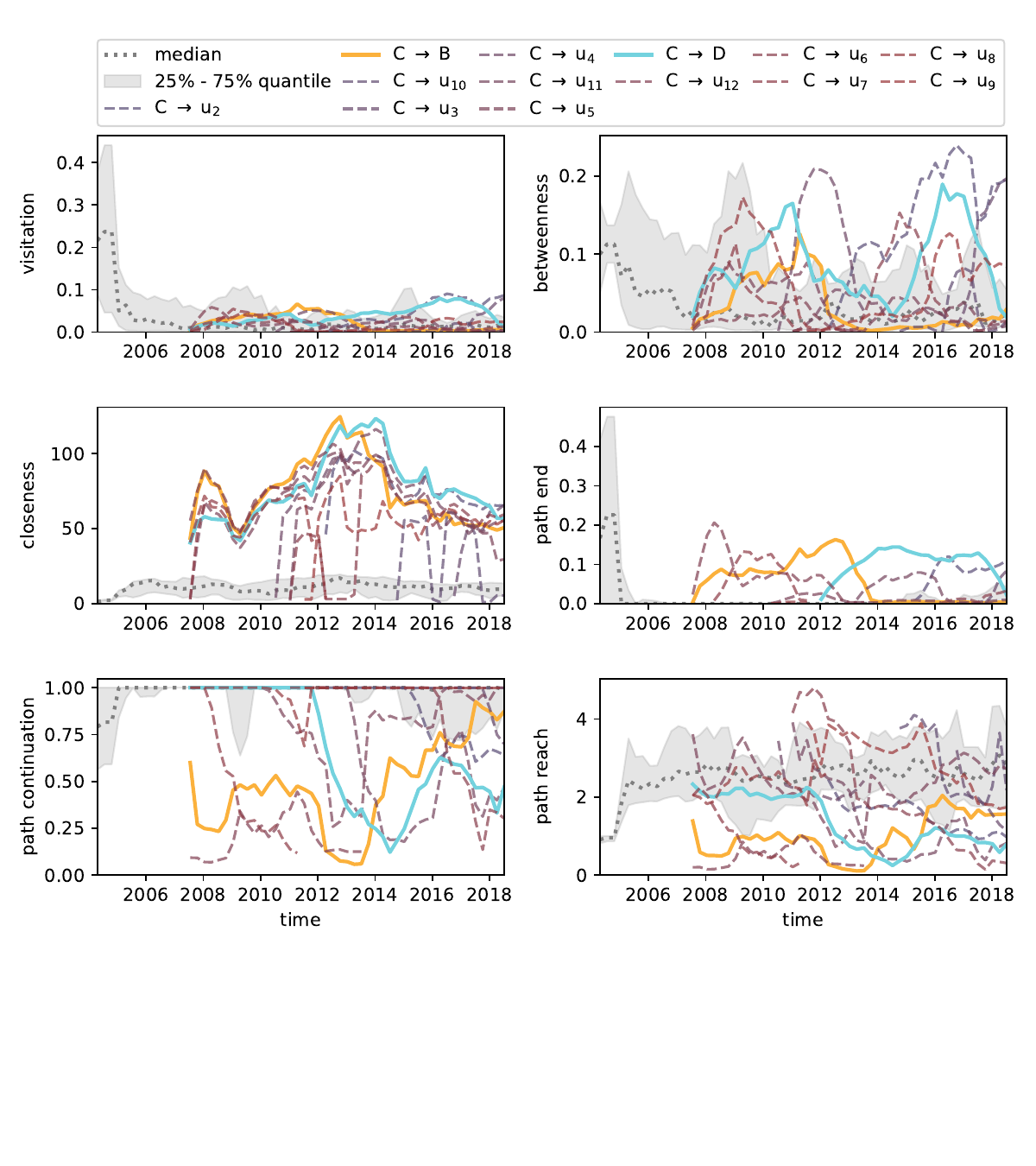}
    \caption{Centrality of all edges in the \texttt{MOGen} model with $K=2$ in the \textit{Aegis} data. All edges from $C$ that reach a total visitation of at least 2\% in any given year are highlighted. The median and the 25\% to 75\% quantile are shown in grey.}
    \label{fig:Centralities_Aegis_2nd_Order_devC_part2}
\end{figure}

\begin{figure}
    \centering
    \includegraphics[width=\textwidth, trim=0 120 0 0, clip]{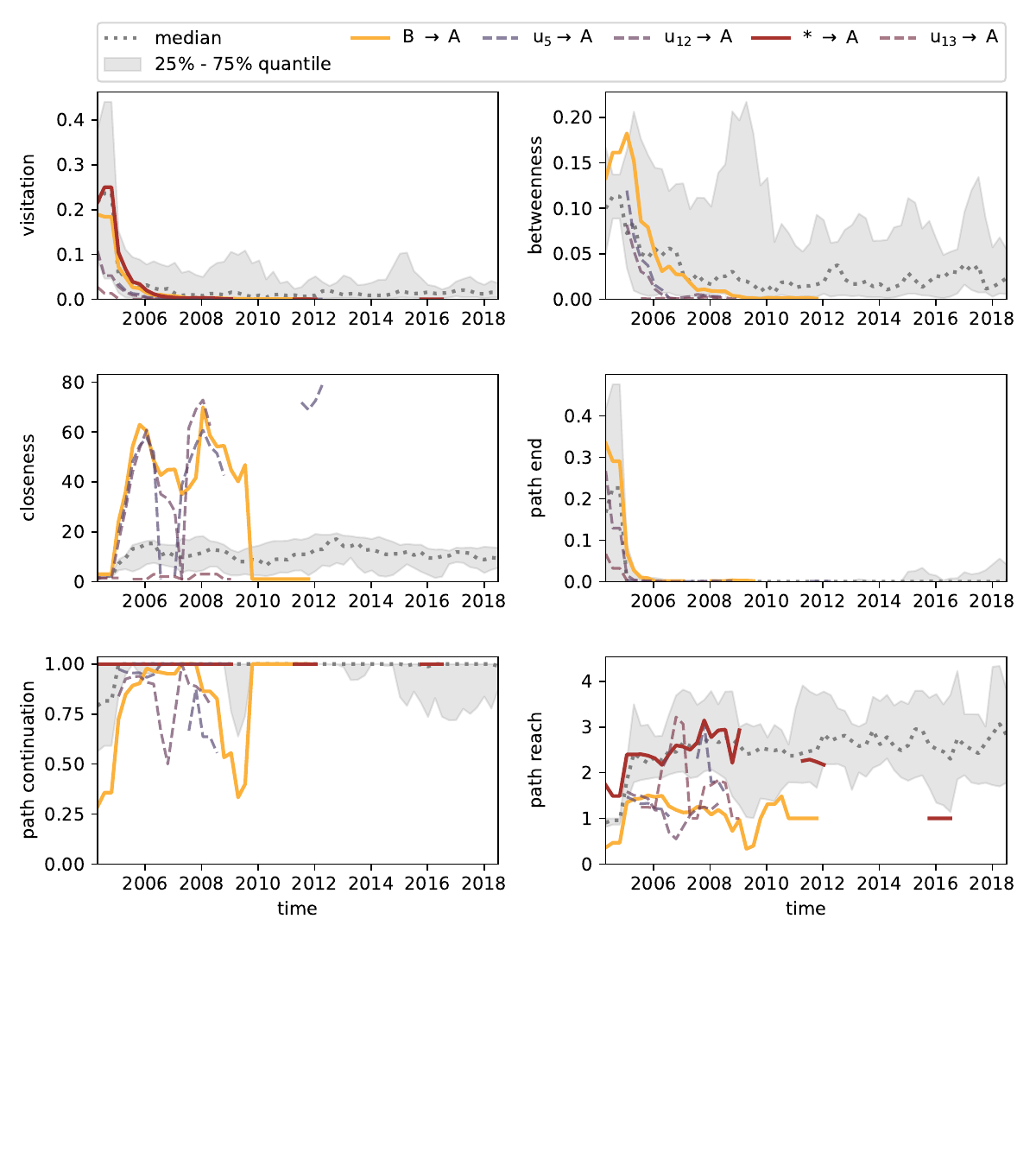}
    \caption{Centrality of all edges in the \texttt{MOGen} model with $K=2$ in the \textit{Aegis} data. All edges to $A$ that reach a total visitation of at least 2\% in any given year are highlighted. The median and the 25\% to 75\% quantile are shown in grey.}
    \label{fig:Centralities_Aegis_2nd_Order_devA}
\end{figure}

\begin{figure}
    \centering
    \includegraphics[width=\textwidth, trim=0 120 0 0, clip]{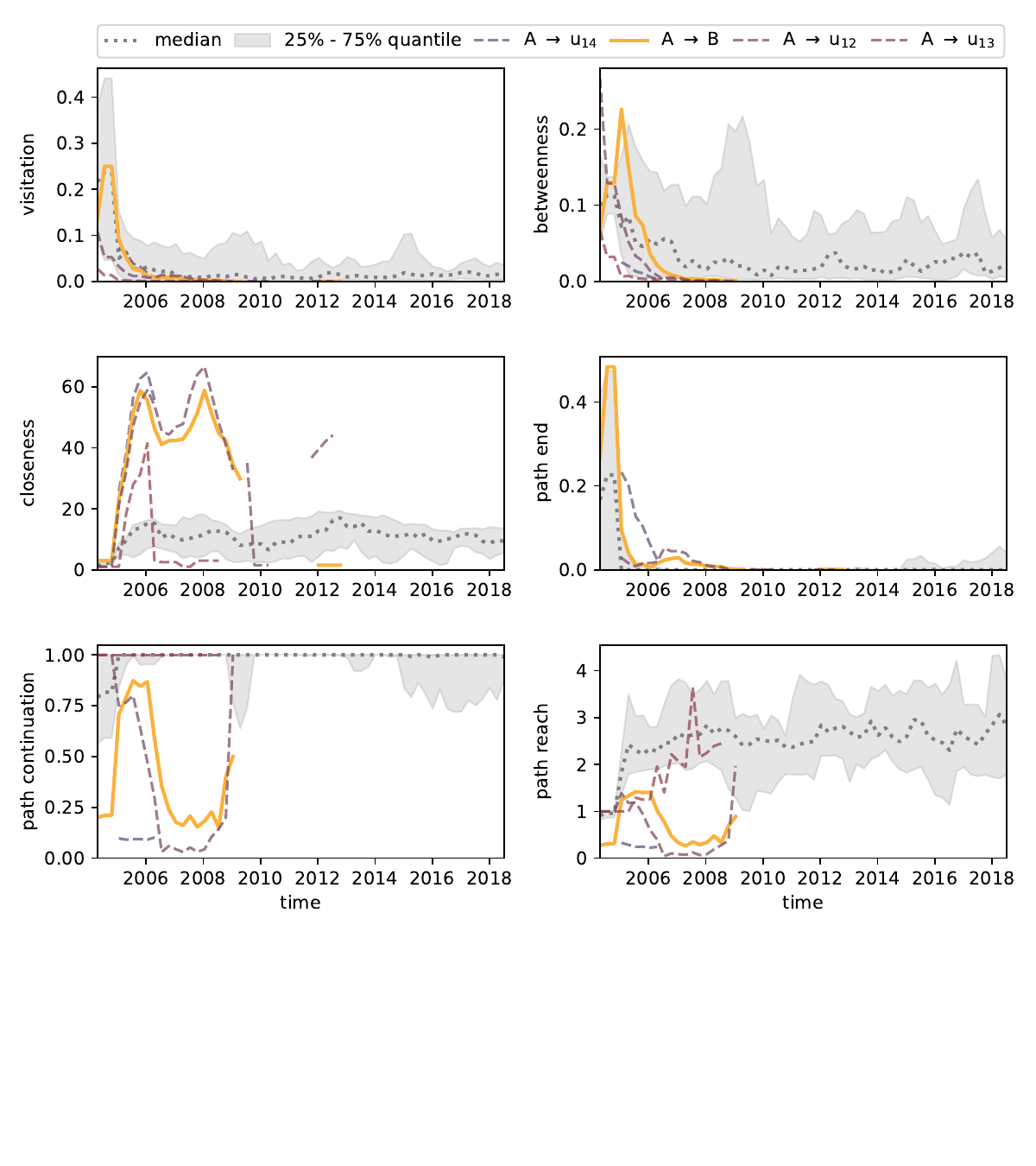}
    \caption{Centrality of all edges in the \texttt{MOGen} model with $K=2$ in the \textit{Aegis} data. All edges from $A$ that reach a total visitation of at least 2\% in any given year are highlighted. The median and the 25\% to 75\% quantile are shown in grey.}
    \label{fig:Centralities_Aegis_2nd_Order_devA_part2}
\end{figure}

\begin{figure}
    \centering
    \includegraphics[width=\textwidth, trim=0 120 0 0, clip]{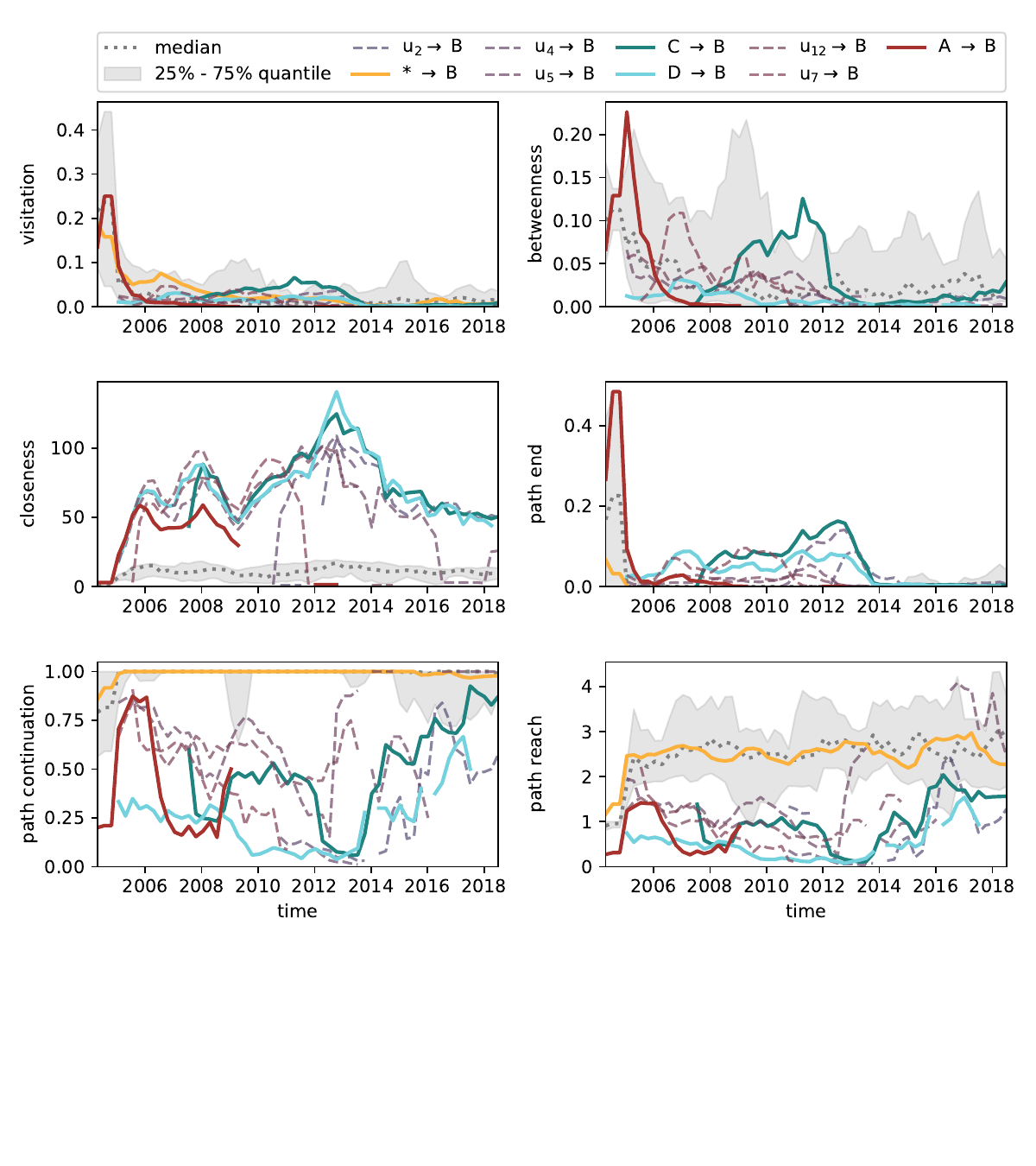}
    \caption{Centrality of all edges in the \texttt{MOGen} model with $K=2$ in the \textit{Aegis} data. All edges to $B$ that reach a total visitation of at least 2\% in any given year are highlighted. The median and the 25\% to 75\% quantile are shown in grey.}
    \label{fig:Centralities_Aegis_2nd_Order_devB}
\end{figure}

\begin{figure}
    \centering
    \includegraphics[width=\textwidth, trim=0 120 0 0, clip]{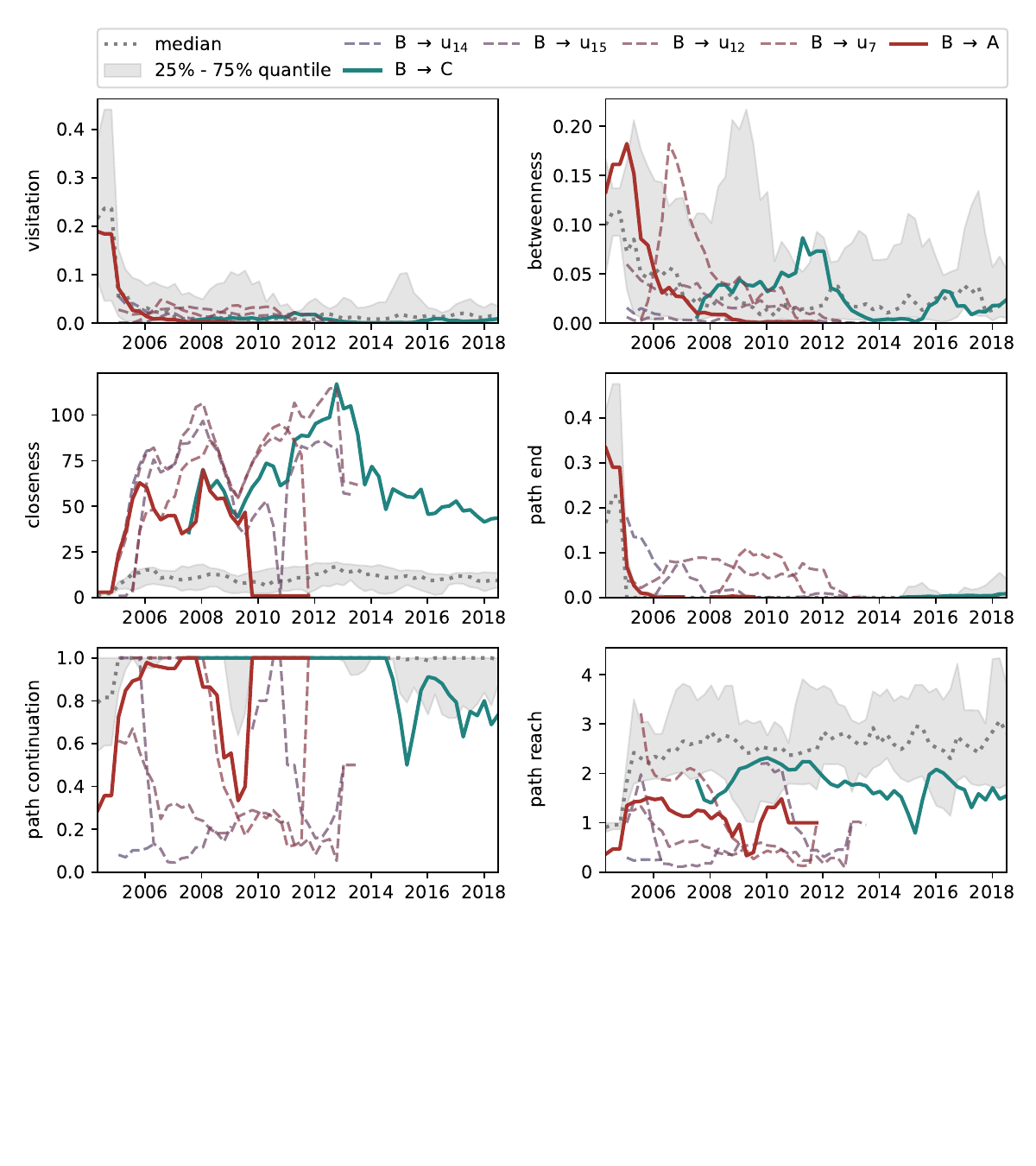}
    \caption{Centrality of all edges in the \texttt{MOGen} model with $K=2$ in the \textit{Aegis} data. All edges from $B$ that reach a total visitation of at least 2\% in any given year are highlighted. The median and the 25\% to 75\% quantile are shown in grey.}
    \label{fig:Centralities_Aegis_2nd_Order_devB_part2}
\end{figure}

\begin{figure}
    \centering
    \includegraphics[width=\textwidth, trim=0 120 0 0, clip]{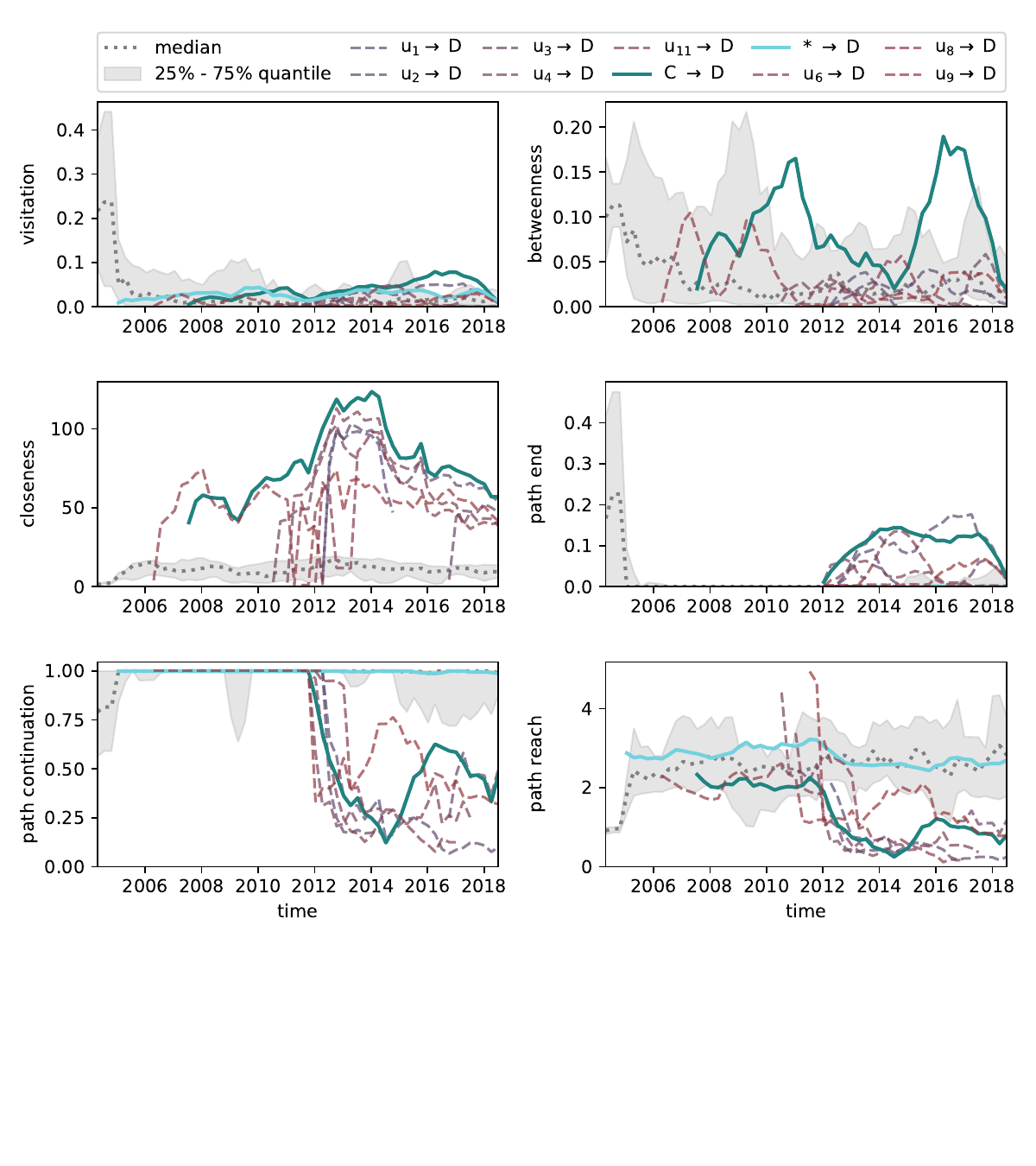}
    \caption{Centrality of all edges in the \texttt{MOGen} model with $K=2$ in the \textit{Aegis} data. All edges to $D$ that reach a total visitation of at least 2\% in any given year are highlighted. The median and the 25\% to 75\% quantile are shown in grey.}
    \label{fig:Centralities_Aegis_2nd_Order_devD}
\end{figure}

\begin{figure}
    \centering
    \includegraphics[width=\textwidth, trim=0 120 0 0, clip]{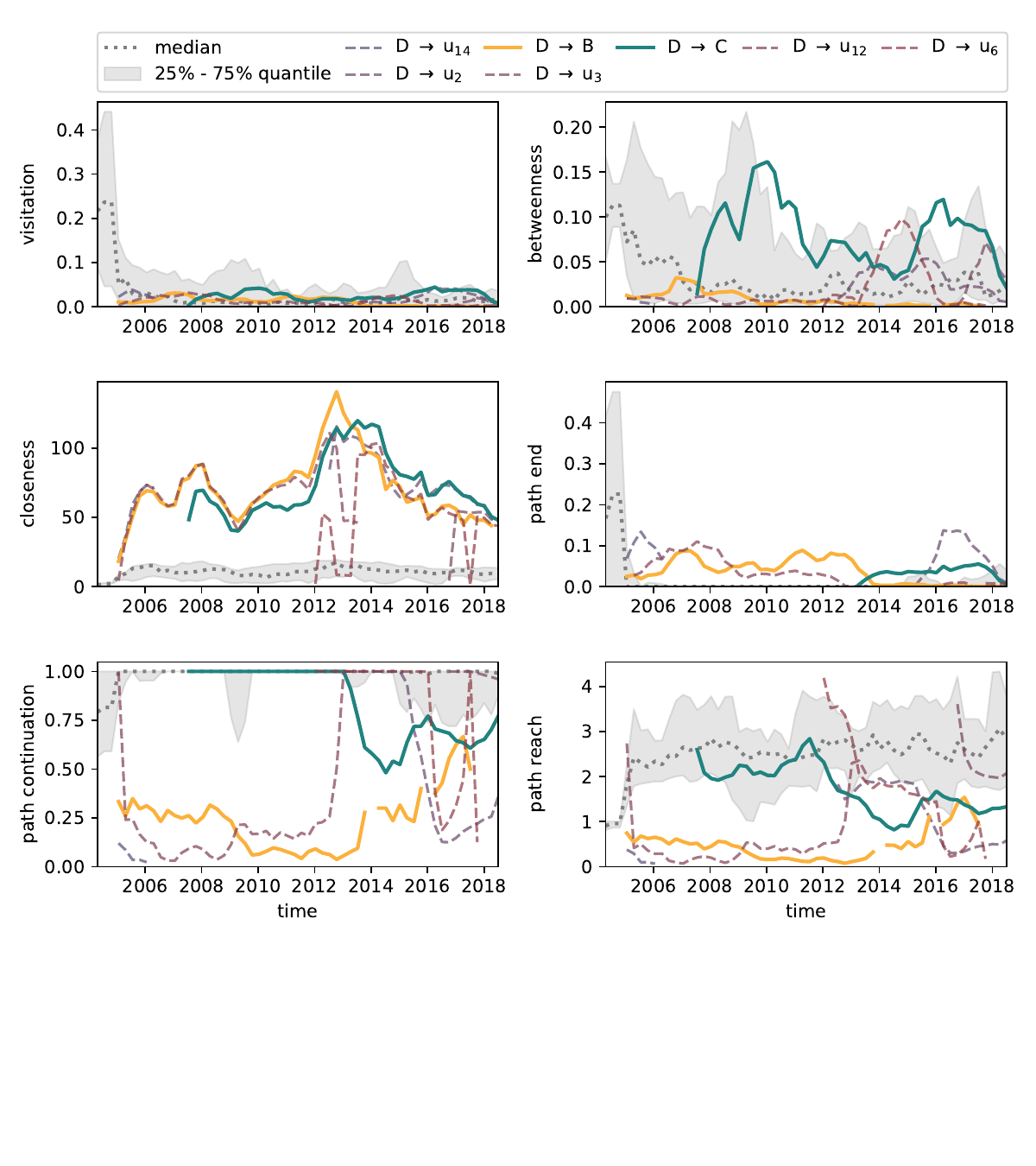}
    \caption{Centrality of all edges in the \texttt{MOGen} model with $K=2$ in the \textit{Aegis} data. All edges from $D$ that reach a total visitation of at least 2\% in any given year are highlighted. The median and the 25\% to 75\% quantile are shown in grey.}
    \label{fig:Centralities_Aegis_2nd_Order_devD_part2}
\end{figure}

\end{appendices}

\end{document}

%% file: MOGen_representation.tex
\begin{figure}
	\centering
	\begin{tikzpicture}[x=.5cm, y=-.5cm]
		\draw[very thick, rounded corners=2] (.5,.5) -| (.3,5) |- (.5, 9.5);
		\draw[very thick, rounded corners=2] (9.5,.5) -| (9.7,5) |- (9.5, 9.5);

		\draw[ultra thick, draw=white, fill=customY!30] (0.5,0.5) rectangle (9.5,9.5);
		\draw[ultra thick, draw=white, fill=black!10] (8.5,0.5) rectangle node {$0$} (9.5,1.5);
		\draw[ultra thick, draw=white, fill=customG!30] (2.5,.5) rectangle node {$0$} (8.5,1.5);
		
		\draw[ultra thick, draw=white, fill=customY!30] (2.5,1.5) rectangle node {\small$\textbf{T}_{1,2}$} (4.5,3.5);
		\draw[ultra thick, draw=white, fill=customY!30] (4.5,3.5) rectangle node[yshift=2] {\small$\vdots$} (5.5,5.5);
		\draw[ultra thick, draw=white, fill=customY!30] (5.5,5.5) rectangle node {\small$\textbf{T}_{K-1,K}$} (8.5,6.5);
		\draw[ultra thick, draw=white, fill=customY!30] (5.5,6.5) rectangle node {\small$\textbf{T}_{K,K}$} (8.5,9.5);
		\node at (3,6.5) {$0$};
		\node at (7,3.5) {$0$};
		
		\draw[ultra thick, draw=white, fill=customG!30] (.5,.5) rectangle node {\small$\textbf{T}_{0,1}$} (2.5,1.5);
		
		\draw[ultra thick, draw=white, fill=customB!30] (8.5,1.5) rectangle node {\small$\textbf{T}_{\dagger}$} (9.5,9.5);
		
		\node[anchor=south] at (1.5,0.5) {\small $V^1$};
		\node[anchor=south] at (3.5,0.5) {\small $V^2$};
		\node[anchor=south] at (5,0.3) {\small $\dots$};
		\node[anchor=south] at (7,0.5) {\small $V^K$};
		\node[anchor=south] at (9,0.5) {\small $\dagger$};
		
		\node at (-.8,1) {\small $*$};
		\node at (-.8,2.5) {\small $V^1$};
		\node at (-.8,4.3) {\small $\vdots$};
		\node at (-.8,6) {\small $V^{K-1}$};
		\node at (-.8,8) {\small $V^K$};
		
		\node[anchor=east] at (-2, 5) {$\textbf{T}^{(K)}=$};
	\end{tikzpicture}
	\caption{Multi-order transition matrix $\textbf{T}^{(K)}$ of a \texttt{MOGen} model with maximum-order $K$. We split $\textbf{T}^{(K)}$ into a transient part $\textbf{Q}$ (\textcolor{customY!30}{\faSquare}) and an absorbing part $\textbf{R}$ (\textcolor{customB!30}{\faSquare}). $\textbf{S}$ (\textcolor{customG!30}{\faSquare}) represents the starting distribution of paths.}\label{fig:MOGen_representation}
\end{figure}

%% file: dataset_description.tex
\begin{table}[h!]
\caption{Summary statistics for our five empirical path datasets.}\label{tab:dataset_statistics}

\centering
\sffamily
\begin{tabularx}{.82\linewidth}{@{}X@{\hspace{-5mm}}rrrrrr@{}}
%

\toprule
{} & \multicolumn{2}{c}{paths} & \multicolumn{2}{c}{nodes on path} & \multicolumn{2}{c}{network topology} \\
\cmidrule(lr){2-3}\cmidrule(lr){4-5}\cmidrule(lr){6-7}
{} &  \multicolumn{1}{c}{total} &  \multicolumn{1}{c}{unique} &  \multicolumn{1}{c}{mean} & \multicolumn{1}{c}{median} & nodes & links \\
\midrule
BMS1  & 59,601 & 18,473 &     2.51 & 1 &            497 &          15,387 \\
TUBE & 4,295,731 & 32,313  &  7.9 & 7 &            276.0 &            663 \\
SCHOOL & 103,260 & 25,831 & 2.5 & 2 &            242 &           8,297 \\
HOSPITAL & 62,676 & 13,578 & 4.8 &            5 &             75 &           1,137 \\
WORK     & 7,832 & 1,170 &  2.5 & 2 &             92 &            753 \\
\bottomrule
\end{tabularx}
\end{table}